\documentclass[a4paper,11pt]{article}
\pdfoutput=1
\usepackage{jheppub}

\usepackage{amssymb,amsmath}
\usepackage[normalem]{ulem}
\usepackage[utf8x]{inputenc}
\usepackage{slashed}
\usepackage{graphicx}
\usepackage{here}
\usepackage{color}
\usepackage{csquotes} 
\usepackage{comment}
\usepackage{mathrsfs}
\usepackage{float}
\usepackage{ascmac}
\usepackage{multirow}
\usepackage{longtable}
\usepackage{bm}

\usepackage[italicdiff]{physics}

\usepackage[hang,small,bf]{caption}
\usepackage[subrefformat=parens]{subcaption}
\newcommand{\TeV}{\mbox{TeV}}
\newcommand{\GeV}{\mbox{GeV}}

\makeatletter
\newcommand*\rel@kern[1]{\kern#1\dimexpr\macc@kerna}
\newcommand*\widebar[1]{%
  \begingroup
  \def\mathaccent##1##2{%
    \rel@kern{0.8}%
    \overline{\rel@kern{-0.8}\macc@nucleus\rel@kern{0.2}}%
    \rel@kern{-0.2}%
  }%
  \macc@depth\@ne
  \let\math@bgroup\@empty \let\math@egroup\macc@set@skewchar
  \mathsurround\z@ \frozen@everymath{\mathgroup\macc@group\relax}%
  \macc@set@skewchar\relax
  \let\mathaccentV\macc@nested@a
  \macc@nested@a\relax111{#1}%
  \endgroup
}
\makeatother

\numberwithin{equation}{section}

\preprint{
\begin{minipage}{5cm}
\small
\flushright
KYUSHU-HET-316
\end{minipage}} 

\title{Diffusion-model approach to flavor models: A case study for $S_4^\prime$ modular flavor model}

\author{Satsuki Nishimura$^{1}$,} 
\author{Hajime Otsuka$^{1,2}$, and} 
\author{Haruki Uchiyama$^{1}$}
\affiliation{
$^1$Department of Physics, Kyushu University, 744 Motooka, Nishi-ku, Fukuoka 819-0395, Japan}
\affiliation{
$^2$Quantum and Spacetime Research Institute (QuaSR), Kyushu University, \\ 744 Motooka, Nishi-ku, Fukuoka 819-0395, Japan}
\emailAdd{nishimura.satsuki@phys.kyushu-u.ac.jp}
\emailAdd{otsuka.hajime@phys.kyushu-u.ac.jp}
\emailAdd{uc.haruki496ym@gmail.com}

\abstract{
We propose a numerical method of searching for parameters with experimental constraints in generic flavor models by utilizing diffusion models, which are classified as a type of generative artificial intelligence (generative AI). 
As a specific example, we consider the $S_4^\prime$ modular flavor model and construct a neural network that reproduces quark masses, the CKM matrix, and the Jarlskog invariant by treating free parameters in the flavor model as generating targets. 
By generating new parameters with the trained network and local optimization, we find various phenomenologically interesting parameter regions.
Additionally, we confirm that the spontaneous CP violation occurs in the $S_4^\prime$ model. 
The diffusion model enables an inverse problem approach, allowing the machine to provide a series of plausible model parameters from given experimental data. 
}

\makeatletter
\gdef\@fpheader{}
\makeatother

\begin{document}

\maketitle

\section{Introduction}

There are many approaches to elucidate the flavor structure of quarks and leptons. 
Among them, the flavor symmetry was utilized in understanding the peculiar patterns of fermion masses and mixings through both bottom-up and top-down approaches. 
A prototypical example of continuous symmetries is a $U(1)$ flavor symmetric model using the Froggatt-Nielsen (FN) mechanism \cite{Froggatt:1978nt}, and non-Abelian discrete symmetries were also broadly employed (see for reviews, Refs.~\cite{Altarelli:2010gt,Ishimori:2010au,Hernandez:2012ra,King:2013eh,King:2014nza,Petcov:2017ggy,Kobayashi:2022moq}). 
When Yukawa couplings transform under the modular symmetry, it belongs to a class of modular flavor symmetric models \cite{Feruglio:2017spp} (see for reviews, Refs.~\cite{Kobayashi:2023zzc,Ding:2023htn}).

\medskip

In most of models with flavor symmetries, there exist free parameters that are not controlled under the flavor symmetries. 
Although the flavor structure of fermions can be controlled by flavor symmetries, a realization of realistic flavor structure requires a breaking of the symmetries by a vacuum expectation value (VEV) of a scalar field (so-called flavon field), charged under the flavor symmetries. 
Since the VEV will also be determined by free parameters of the model, these parameters play an important role in evaluating fermion masses and mixing angles quantitatively.  
So far, we have often adopted a certain optimization method, such as the Monte-Carlo simulation, to address the flavor structure of quarks and leptons. 
When we search model parameters using the traditional optimization methods, obtained results are sensitive to the initial values of the model parameters, indicating that it will be difficult to find out realistic flavor patterns from a broad theoretical landscape in a short time. 

\medskip

To achieve highly efficient learning, a machine learning approach is essential.
For instance, reinforcement learning was used to address the flavor structure of quarks and leptons in the $U(1)$ FN model \cite{Harvey:2021oue,Nishimura:2023nre,Nishimura:2024apb}. 
Recently, a diffusion model known as one of generative artificial intelligence (generative AI), was utilized to explore the unknown flavor structure of neutrinos in the context of the Standard Model with three right-handed neutrinos \cite{Nishimura:2025rsk}. 
In particular, the conditional diffusion model was employed to search for model parameters. Although the diffusion model was often applied in generating images where the data of various paintings is collected as input data $G$ with a certain associated information $L$, Ref.~\cite{Nishimura:2025rsk} proposed that the input data $G$ corresponds to a set of model parameters and the label $L$ is specified as neutrino masses and mixing angles. 
Then, the conditional diffusion model successfully reproduces the neutrino mass-squared differences and mixing angles with current experimental constraints, and exhibits non-trivial distributions of the leptonic CP phases and the sums of neutrino masses. 

\medskip

In this paper, we provide a framework for applying the conditional diffusion model to flavor models, where the input data $G$ corresponds to model parameters and $L$ is specified as physical observables such as fermion masses and mixing angles. 
By randomly preparing the model parameters, the diffusion model learns to predict noise in a diffusion process and generates new data in an inverse process in the context of the conditional label $L$. 
As a concrete example, we analyze a specific flavor model, i.e., $S_4^\prime$ modular flavor model, to address the flavor structure of quarks. 
Since the flavor structure of quarks is highly dependent on the value of the symmetry breaking field (modulus $\tau$), the usual optimization methods are much more sensitive to initial values in numerical simulations. 
Hence, a limited region in the parameter space has been explored in e.g., Refs.~\cite{Novichkov:2020eep,Liu:2020akv,Abe:2023qmr}. By applying the conditional diffusion model to model parameters in the $S_4^\prime$ modular flavor model, it turns out that a semi-realistic flavor structure of quarks can be realized at a certain value of the modulus $\tau$ in a short time. Then, the model parameters can be accurately fitted to the experimental data through local optimization methods such as Markov Chain Monte-Carlo (MCMC).
Note that the obtained parameter region is different from that in the previous literature. 
Furthermore, the CP symmetry will be spontaneously broken by the modulus $\tau$. 
Therefore, our proposed diffusion-model approach will be regarded as an alternative numerical method to address the flavor structure of quarks. 

\medskip

This paper is organized as follows. 
The conditional diffusion model is introduced for flavor models in Sec.~\ref{sec:DM_flavor}, and Sec.~\ref{sec:S4prime_model} presents the $S_4^\prime$ modular flavor model as a practical application. 
In that section, we describe the modular symmetry in Sec.~\ref{sec:S4prime_modular}. 
Based on this symmetry, the quark sector of the $S_4^\prime$ modular flavor model is organized in Sec.~\ref{sec:S4prime_quark}, and we construct a concrete diffusion model in Sec.~\ref{sec:diffusion_model}. 
We discuss the results generated by the diffusion model in Sec.~\ref{sec:result}. 
Finally, Sec.~\ref{sec:con} is devoted to the conclusion and future prospects.

\section{Diffusion models for flavor physics}
\label{sec:DM_flavor}

In this section, we provide a brief introduction to denoising diffusion probabilistic models (DDPMs) \cite{Ho:2020epu} with classifier-free guidance (CFG) \cite{ho:2022cla}. 
They provide an intuitive definition of conditional diffusion models. 
For further details regarding the formulation of DDPMs and CFG, see the Appendices of Ref.~\cite{Nishimura:2025rsk}.

\newpage

\medskip

The diffusion model consists of two stages: (i) the diffusion process and (ii) the inverse process. 
In the diffusion process, noise is added to the input data and a machine learns to predict the added noise. 
In the inverse process, noise is gradually removed from a pure noise to generate meaningful data. 
In the context of flavor models, let us denote $G$ as free parameters in flavor models and $L$ as a conditional label specifying physical observables such as masses and mixings. 
For instance, the lepton sector was analyzed by incorporating the Yukawa couplings into $G$ and the neutrino masses and the PMNS matrix into $L$ in Ref.~\cite{Nishimura:2025rsk}. 
In this study, we aim to apply this method to the quark sector.

\medskip

Based on an initial input data $x_{0}=G$, a series of new data $\{x_{1},x_{2},\ldots,x_{T}\}$ is defined as Markov process that adds noise to the input $x_0$:
\begin{align}
    q\left(x_{1:T}|x_{0}\right) &= \prod_{t=1}^{T} q\left(x_{t}|x_{t-1}\right), \\
    q\left(x_{t}|x_{t-1}\right) &= \mathcal{N} \left(x_{t}; \sqrt{A_{t}}x_{t-1}, B_{t}\right),
\end{align}
with $x_{i:j} = x_{i}, x_{i+1}, \ldots, x_{j}$.
Here, a conditional probability $q\left(x_{t}|x_{t-1}\right)$ denotes the probability of $x_{t}$ given the condition $x_{t-1}$.
In addition, $\mathcal{N} \left(x; \mu, \sigma\right)$ is a normal distribution with a random variable $x$, an average $\mu$, and a variance $\sigma$.
We will omit $x$ for brevity.
$A_t$ is defined as $A_{t}=1-B_{t}$, and the parameters $0<B_{1}<B_{2}<\cdots<B_{T}<1$ decide the variance of $\mathcal{N}$.
By the definition of $A$ and $B$, the series of data close to pure noise that follows a standard normal distribution along with $t$. 
Based on this property, $A_t$ and $B_t$ are referred to as noise schedules. 

\medskip

For practical purposes, the noised data at any given time is determined as
\begin{align}
    x_{t} &= \sqrt{\bar{A}_t} x_0 + \sqrt{\bar{B}_t} \epsilon, \\
    \bar{A}_{t} &= \prod_{s=1}^{t} A_{s}, \quad
    \bar{B}_{t} = 1 - \bar{A}_{t},
    \label{eq:data_sequence}
\end{align}
with $t=1,\ldots,T$. 
Here, $\epsilon$ is a noise obeying a standard normal distribution $\mathcal{N} \left(0, 1\right)$.
Among the various ways to choose $B_t$, we adopt a linear schedule:
\begin{align}
    B_{t} = \left(1-\frac{t}{T}\right) B_{\mathrm{min}} + \frac{t}{T} B_{\mathrm{max}}.
\end{align}
We adopt $B_{\mathrm{min}}=10^{-4}$, $B_{\mathrm{max}}=0.02$, and $T=1000$.

\medskip

Then, in the reverse process, pure noise $x_{T}$ from $\mathcal{N} \left(0, 1\right)$ is prepared as initial data, and $x_{t-1}$ based on $x_{t}$ is sampled in sequence according to the following Markov process:
\begin{align}
    p_{\theta} \left(x_{0:T}\right) &= p\left(x_{T}\right) \prod_{t=1}^{T} p_{\theta} \left(x_{t-1}|x_{t}\right), \\
    p\left(x_{T}\right) &= \mathcal{N} \left(x_{T}; 0, 1\right).
\end{align}
The conditional probability $p_{\theta} \left(x_{t-1}|x_{t}\right)$ at each step is estimated using a neural network, which is explained later, and characterized by parameters $\theta$. 
In practice, sampling from $p_{\theta}$ corresponds to determining $x_{t-1}$ as follows:
\begin{align}
    &x_{t-1} = \frac{1}{\sqrt{A_t}}\left( x_t - \frac{B_t}{\sqrt{\bar{B}_t}}\hat{\epsilon}_\theta \right) + \sigma_{t} u_{t}, \label{eq:denoise} \\
    &\hat{\epsilon}_\theta\left(x_t,t,L\right) = \left(1-\gamma\right)\epsilon_\theta\left(x_t,t\right) + \gamma \epsilon_\theta\left(x_t,t,L\right), \label{eq:CFG_lin_comb}
\end{align}
where $L$ is a conditional label and $u_{t}$ is a disturbance following $\mathcal{N} \left(0, 1\right)$.
In addition, the coefficient $\gamma$, which is called the CFG scale, satisfies the condition $\gamma \geq 0$. 
A larger CFG scale means that the data are more faithful to the label $L$, but the diversity of generated results tends to be lost. 
On the other hand, when $\gamma<1$, it emphasizes the diversity.
$\gamma = 8$ is adopted in this study, and this value is often used in image generation also.

\medskip

The predicted noise $\hat{\epsilon}_\theta$ is determined by a linear combination of the unlabeled noise and the labeled noise. 
In contrast to the expression of Eq.~\eqref{eq:CFG_lin_comb}, just one mathematical model is required by considering $\epsilon_\theta(x_t,t) = \epsilon_\theta(x_t,t,L=\varnothing)$. 
For practical training, $L=\varnothing$ is adopted with a low probability (10-20\%), and the learning of CFG proceeds with mixing the cases with and without labels. 
For $\varnothing$, a learned embedding vector or a zero vector $\varnothing=0$ is often used.
In this study, we dropped out the labels in 10\% of the cases, and 0 is used as $\varnothing$ in our study.

\medskip

\begin{figure}[H]
    \centering
    \includegraphics[width=140mm]{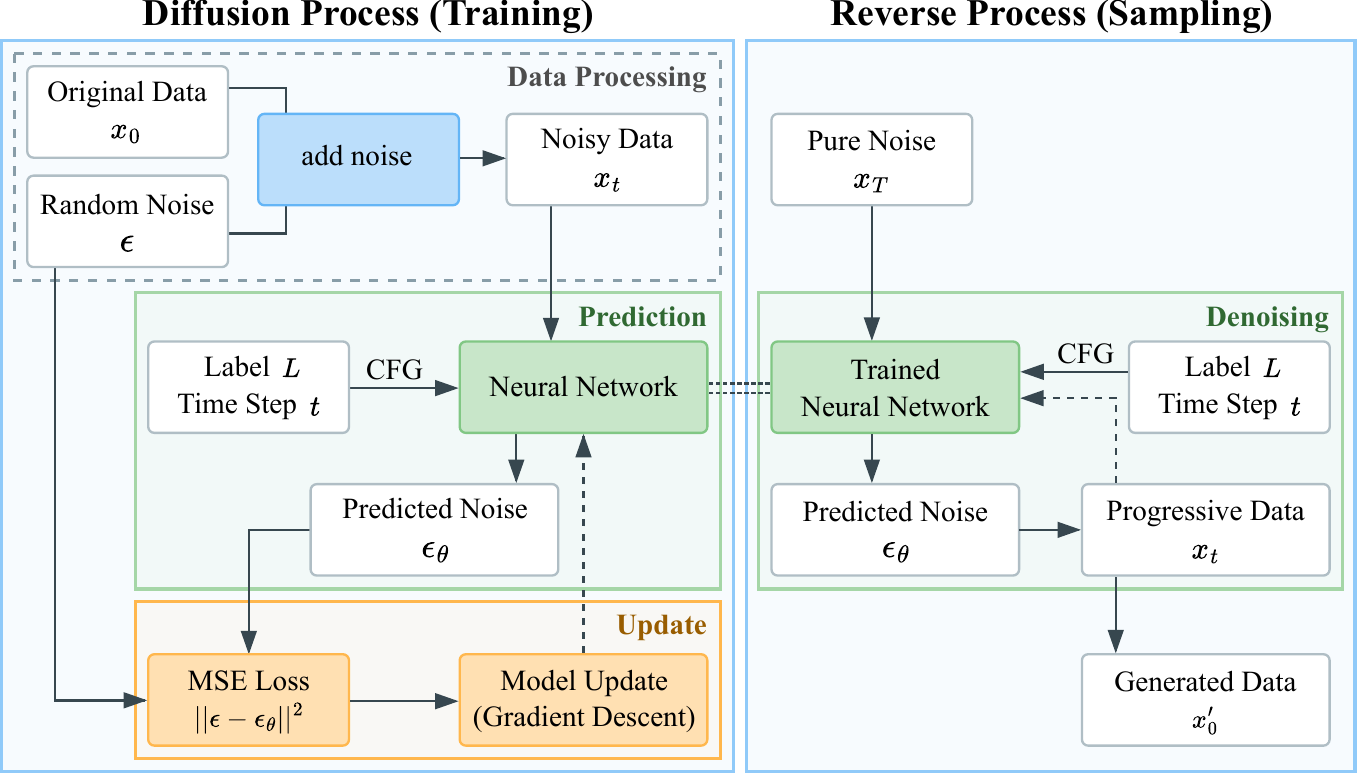}
    \caption{The data flow in DDPM with CFG. The neural network predicts an added noise based on the noised data and conditional labels $\{L,t\}$. The parameter $\theta=\{w,b\}$ of that network is updated to minimize the difference between the actual noise $\epsilon$ and the predicted noise $\epsilon_\theta$.}
    \label{fig:diffusion_process}
\end{figure}

To predict the added noise, we utilize a neural network model in which a $n$-th layer with $N_{n-1}$ dimensional vector $\Vec{X}_{n-1} = (X_{n-1}^{1},\ldots, X_{n-1}^{N_{n-1}})$ transforms into a $N_n$ dimensional vector $\Vec{X}_n = (X_{n}^{1},\ldots, X_{n}^{N_{n}})$ by following relation: 
\begin{align}
X_{n}^{i} = h_n \left( w^{ij}_n X_{n-1}^{j} + b^i_n \right), \label{eq:activation}
\end{align}
where $h,w,b$ respectively denote the activation function, the weight and the bias.
In general, the neural network realizes multiple nonlinear transformations through $h$. 
In our architecture, a fully-connected network is adopted, and the network is trained by minimizing the mean squared error (MSE) loss representing the difference between the actual added noise $\epsilon$ and the predicted noise $\epsilon_\theta$, as shown in Fig.~\ref{fig:diffusion_process}. 
In the end, a well-trained network can precisely estimate the noise component in $x_{t}$ as the output of the calculation shown in Eq.~\eqref{eq:activation}.
Thus, during data generation, the new final data $x_{0}$ is obtained by repeatedly applying the denoising process according to Eq.~\eqref{eq:denoise}.

\medskip

When we complete training a neural network once using initial data, the inverse process with that network can provide various new data $G$ including free parameters in flavor models. 
If the network does not reach the desired level of accuracy, many types of strategies are possible to tackle this problem.
To enhance learning efficiency and accuracy, transfer learning is often applied by reusing a neural network that has already been learned. 
In a more narrowly defined method called fine-tuning, the parameters of the learned neural network are used as the initial values of a new neural network, and all parameters are updated based on other training data. 
This study also implements this method and evaluates its effectiveness.
The details of transfer learning and fine-tuning are provided in Ref.~\cite{Nishimura:2025rsk}.

\section{$S_4^\prime$ modular flavor model as a case study}
\label{sec:S4prime_model}

In this section, we apply the conditional diffusion model to a particular flavor model, namely the modular flavor model. 
By introducing the modular symmetry in Sec.~\ref{sec:S4prime_modular}, we present the $S_4^\prime$ modular flavor model in Sec.~\ref{sec:S4prime_quark}. 
The formulation of the conditional diffusion model is discussed in Sec.~\ref{sec:diffusion_model}.

\subsection{Modular symmetry}
\label{sec:S4prime_modular}

In this section, we review the $SL(2,\mathbb{Z})$ modular symmetry and introduce the $S_4^\prime$ modular flavor model. 
A principal congruence subgroup $\Gamma(N)$ of $SL(2,\mathbb{Z})$ is defined as follows:
\begin{align}
    \Gamma(N) = \left\{
    \begin{pmatrix} a & b \\ c & d \end{pmatrix} 
    \in SL(2,\mathbb{Z}),\quad
    \begin{pmatrix} a & b \\ c & d \end{pmatrix} \equiv 
    \begin{pmatrix} 1 & 0 \\ 0 & 1 \end{pmatrix}\,\mathrm{mod}\,N
    \right\}.
\end{align}
This group acts on the complex variable $\tau$ called modulus ($\Im\,[\tau]>0$) in the following way:
\begin{align}
 \tau \to \frac{a\tau +b}{c\tau+d},    
\end{align}
with $ad-bc=1$.
In addition, $\Gamma(N)$ is generated by the three generators:
\begin{align}
    S = \begin{pmatrix} 0 & 1 \\ -1 & 0 \end{pmatrix} , \qquad 
    T = \begin{pmatrix} 1 & 1 \\ 0 & 1 \end{pmatrix} , \qquad
    R = \begin{pmatrix} -1 & 0 \\ 0 & -1 \end{pmatrix}.
\end{align}
These generators satisfy the following algebraic relations:
\begin{align}
    S^2 = R, \qquad 
    (ST)^3 = R^2 = \mathbb{I}, \qquad 
    TR = RT.
\end{align}

\medskip

Using the definition of $\Gamma(N)$, various finite modular groups are defined as follows:
\begin{align}
    PSL(2,\mathbb{Z}) &= SL(2,\mathbb{Z}) / \mathbb{Z}_2^R, \\
    \Gamma_N^\prime &= SL(2,\mathbb{Z}) / \Gamma(N), \\
    \Gamma_N &= PSL(2,\mathbb{Z}) / \Gamma(N).
\end{align}
Note that $\Gamma_N$ with $N=2,3,4,5$ are isomorphic to $S_3, A_4, S_4, A_5$, respectively.
Similarly, $\Gamma_N^\prime$ with $N=3,4,5$ are isomorphic to $A_4^\prime, S_4^\prime, A_5^\prime$, which are double covering groups of $A_4, S_4, A_5$, respectively.
In these groups, the generator $T$ satisfies $T^N=\mathbb{I}$, so it generates a $\mathbb{Z}_N^T$ symmetry. 

\medskip

In this paper, we focus on $S_4^\prime$, having ten irreducible representations:
\begin{align}
    \mathbf{1},\mathbf{1}^\prime, \mathbf{2}, \mathbf{3}, \mathbf{3}^\prime \quad
    \mathrm{and} \quad
    \hat{\mathbf{1}},\hat{\mathbf{1}}^\prime, \hat{\mathbf{2}}, \hat{\mathbf{3}}, \hat{\mathbf{3}}^\prime.
\end{align}
The non-hatted and hatted representations respectively transform under $R$ trivially and non-trivially. 
In other words, $Rr = r$ and $R\hat{r} = -\hat{r}$ are satisfied for any representation $r$ and $\hat{r}$. 
In this work, we choose representation matrices in which $T$ is diagonal and $S$ is real. 
For the doublet $\mathbf{2}$ and the triplet $\mathbf{3}$, the representation matrices is chosen as follows \cite{Novichkov:2020eep}:
\begin{align}
    \rho_S(\mathbf{2}) = \frac{1}{2} \begin{pmatrix}
    -1 & \sqrt{3} \\ \sqrt{3} & 1 
    \end{pmatrix}, &\quad 
    \rho_T(\mathbf{2}) = \begin{pmatrix}
    1 & 0 \\ 0 & -1 
    \end{pmatrix}, \\
    \rho_S(\mathbf{3}) = -\frac{1}{2} \begin{pmatrix}
    0 & \sqrt{2} & \sqrt{2} \\ 
    \sqrt{2} & -1 & 1 \\ 
    \sqrt{2} & 1 & -1 \\ 
    \end{pmatrix}, &\quad 
    \rho_T(\mathbf{3})= 
    \begin{pmatrix}
    -1 & 0 & 0 \\ 0 & -i & 0 \\ 0 & 0 & i 
    \end{pmatrix}. 
\end{align}
Then, the primed/hatted representations obey following relations for any representation $r$ and $\hat{r}$: 
\begin{align}
    \rho_S(r) &=-\rho_S(r^\prime) = -i \rho_S(\hat{r}) = i \rho_S(\hat{r}^\prime), \\ 
    \rho_T(r) &=-\rho_T(r^\prime) = i \rho_T(\hat{r}) = -i \rho_T(\hat{r}^\prime), \\
    \rho_R(r) &=\rho_R(r^\prime) = - \rho_R(\hat{r}) = -\rho_R(\hat{r}^\prime) = 1.
\end{align}

\medskip

In the upper half-plane in the complex plane of $\tau$, a modular form $Y_r^{(k)}$ of representation $r$ with a weight $k$ is defined as a holomorphic function which transforms in the following rule under the modular symmetry:
\begin{align}
    Y_{r}^{(k)}(\tau) \to (c\tau+d)^{k} \rho(r) Y_r^{(k)}(\tau),
\end{align}
with a representation matrix $\rho(r)$. 
In addition, a matter field $\phi$ is assumed to obey the same rule as follows:
\begin{align}
    \phi \to  (c\tau+d)^{k_\phi} \rho(r_\phi) \phi,
\end{align}
with representation $r_\phi$ and weight $k_\phi$.

\medskip

To present an explicit expression of the modular form under the $S_4^\prime$ modular symmetry, let us introduce the Dedekind eta function $\eta(\tau)$:
\begin{align}
    \eta (\tau) = e^{\pi i \tau/12} \prod_{n=1}^{\infty} \left( 1-e^{2\pi i \tau n} \right).
\end{align}
Using this, two functions $\theta$ and $\epsilon$ are defined as follows:
\begin{align}
\begin{split}
    \theta (\tau) &= \frac{\eta(2\tau)^5}{\eta(\tau)^2 \eta(4\tau)^2} = 1 + 2 \sum_{n=1}^{\infty} q^{n^2}, \\
    \epsilon (\tau) &= \frac{2\eta(4\tau)^5}{\eta(2\tau)}= 2q^{1/4} \sum_{n=0}^{\infty} q^{n(n+1)},
\end{split}
\end{align}
where we show their $q$-expansions with $q=e^{2\pi i \tau}$. 
Under the $S_4^\prime$ modular symmetry, there is a $\hat{\mathbf{3}}$ representation with $k_Y=1$, which is referred to in Ref.~\cite{Novichkov:2020eep}:
\begin{align}
    Y^{(1)}_{\hat{\mathbf{3}}}(\tau) = \begin{pmatrix}
    \sqrt{2}\epsilon(\tau) \theta(\tau) \\ \epsilon^2(\tau) \\ -\theta^2(\tau)
    \end{pmatrix}.
\end{align}
Modular forms with higher weight are calculated by tensor products given in Ref.~\cite{Novichkov:2020eep}. 
In the following, we list relevant modular forms utilized in this work:
\begin{align}
\begin{split}
    Y_\mathbf{3}^{(4)} &= \epsilon \theta (\epsilon^4 - \theta^4)
    \left(
    \begin{array}{c}
         -\sqrt{2}\epsilon \theta  \\
         -\epsilon^2\\
         \theta^2
    \end{array}
    \right),\quad
    Y_\mathbf{3}^{(6)} = \epsilon \theta (\epsilon^4 - \theta^4)
    \left(
    \begin{array}{c}
         -2\sqrt{2}\epsilon \theta  (\epsilon^4 + \theta^4)\\
         \epsilon^2 (\epsilon^4 - 5\theta^4)\\
         \theta^2 (5\epsilon^4 - \theta^4)
    \end{array}
    \right), \\
    Y_\mathbf{3}^{1,(8)} &= \epsilon \theta (\epsilon^4 - \theta^4)
    \left(
    \begin{array}{c}
         16\sqrt{2}\epsilon^5 \theta^5  \\
         \epsilon^2 (\epsilon^8 +10\epsilon^4 \theta^4 + 5\theta^8)\\
         -\theta^2 (5\epsilon^8 +10\epsilon^4 \theta^4 + \theta^8)
    \end{array}
    \right), \\
    Y_\mathbf{3}^{2,(8)} &= \epsilon^2 \theta^2 (\epsilon^4 - \theta^4)^2
    \left(
    \begin{array}{c}
         \epsilon^4 - \theta^4\\
         2\sqrt{2}\epsilon \theta^3\\
         2\sqrt{2}\epsilon^3 \theta
    \end{array}
    \right), \\
    Y_{\hat{\mathbf{3}}}^{1,(7)} &= 
    \left(
    \begin{array}{c}
         32\sqrt{2}\epsilon^5 \theta^5 (\epsilon^4 + \theta^4) \\
         -\epsilon^2 (\epsilon^{12} -19\epsilon^8 \theta^4 - 45\epsilon^4\theta^8 -\theta^{12})\\
         -\theta^2 (\epsilon^{12} +45\epsilon^8 \theta^4 +19\epsilon^4\theta^8 -\theta^{12})
    \end{array}
    \right), \\
    Y_{\hat{\mathbf{3}}}^{2,(7)} &= \epsilon \theta (\epsilon^4 - \theta^4)
    \left(
    \begin{array}{c}
         \epsilon^8 - \theta^8\\
         -\sqrt{2}\epsilon \theta^3  (7\epsilon^4 + \theta^4)\\
         -\sqrt{2}\epsilon^3 \theta (\epsilon^4 + 7\theta^4)
    \end{array}
    \right).
\end{split}
\end{align}

\subsection{Quark sector}
\label{sec:S4prime_quark}

Then, we consider a specific example of $S_4^\prime$ modular flavor model with an emphasis on the quark sector. Following Ref.~\cite{Abe:2023qmr}, we suppose that 
$Q$ is a triplet representation of $S_4^\prime$, $\{u_1^c, u_2^c, d_1^c, d_2^c, d_3^c\}$ are trivial singlets of $S_4^\prime$, and $u_3^c$ are the non-trivial singlets of $S_4^\prime$, i.e.,
\begin{align}
    Q\,:\,\mathbf{3},\qquad
    u^c\,:\,\mathbf{1}\oplus \mathbf{1} \oplus \hat{\mathbf{1}}^\prime,\qquad    
    d^c\,:\,\mathbf{1}\oplus \mathbf{1} \oplus \mathbf{1}.
\end{align}
In addition, Higgs doublets $H_u$ and $H_d$ are considered trivial singlets $\mathbf{1}$. 
These assignments are summarized in Table \ref{tab:rep_weight_q}.
In the previous study, it was investigated that the CKM matrix has diagonal textures by this configuration.

\begin{table}[H]
\centering
\begin{tabular}{|c||c|ccc|ccc|cc|}
    \hline
     & $Q$ & $u_1^c$ & $u_2^c$ & $u_3^c$ & $d_1^c$ & $d_2^c$ & $d_3^c$ & $H_u$ & $H_d$ \\
    \hline
    $S_4^\prime$ & $\mathbf{3}$ & $\mathbf{1}$ & $\mathbf{1}$ & $\hat{\mathbf{1}}^\prime$ & $\mathbf{1}$ & $\mathbf{1}$ & $\mathbf{1}$ & $\mathbf{1}$ & $\mathbf{1}$ \\
    Weight & $-4$ & $0$ & $-4$ & $-3$ & $0$ & $-2$ & $-4$ & 0 & 0 \\
    \hline
\end{tabular}
\caption{Assignments of representation for the quarks and Higgs doublets and modular weights.}
\label{tab:rep_weight_q}
\end{table}

\medskip

Under this setup, the superpotential in the quark sector is given by
\begin{align}
    W_{q} = &H_u \left\{ \alpha_1 (Q Y_\mathbf{3}^{(4)})_\mathbf{1} u_1^c + \sum_{a=1}^{2} \left[ \alpha_2^a (Q Y_\mathbf{3}^{a,(8)})_\mathbf{1} u_2^c + \alpha_3^a ((Q Y_{\hat{\mathbf{3}}}^{a,(7)})_{\hat{\mathbf{1}}} u_3^c)_\mathbf{1} \right]
    \right\}
    \nonumber\\
    &+ H_d \left\{ \beta_1 (Q Y_\mathbf{3}^{(4)})_\mathbf{1} d_1^c + \beta_2 (Q Y_\mathbf{3}^{(6)})_\mathbf{1} d_2^c + \sum_{a=1}^{2} \beta_3^a (Q Y_\mathbf{3}^{a,(8)})_\mathbf{1} d_3^c
    \right\}.
\end{align}
In each term, $(\cdots)_{\mathbf{1}}$ is a trivial singlet among the results of the tensor product in the parentheses.
The products of the triplet representations are given by
\begin{align}
    (\mathbf{3}(\phi) \otimes \mathbf{3}(\psi))_1 
    &= \phi_1 \psi_1 + \phi_2 \psi_3 + \phi_3 \psi_2, \\
    (\mathbf{3}(\phi) \otimes \hat{\mathbf{3}}(\psi))_{\hat{1}} 
    &= \phi_1 \psi_1 + \phi_2 \psi_3 + \phi_3 \psi_2.
\end{align}
Then, mass terms of the up-type quarks are derived as follows:
\begin{align}
    &\langle H_u\rangle \left\{ \alpha_1 \left(Q_1 (Y_\mathbf{3}^{(4)})_1 +Q_2 (Y_\mathbf{3}^{(4)})_3 + Q_3 (Y_\mathbf{3}^{(4)})_2\right) u_1^c    \right\} 
   \nonumber\\
   &+\langle H_u\rangle \left\{ \sum_{a=1}^{2}  \alpha_2^a \left(Q_1 (Y_\mathbf{3}^{a,(8)})_1 + Q_2 (Y_\mathbf{3}^{a,(8)})_3 + Q_3 (Y_\mathbf{3}^{a,(8)})_2 \right) \right\}u_2^c
    \nonumber\\
   &+\langle H_u\rangle \left\{ \sum_{a=1}^{2} \alpha_3^a \left(Q_1 (Y_{\hat{\mathbf{3}}}^{a,(7)})_1 + Q_2 (Y_{\hat{\mathbf{3}}}^{a,(7)})_3 + Q_3 (Y_{\hat{\mathbf{3}}}^{a,(7)})_2 \right) \right\}u_3^c,
\end{align}
i.e.,
\begin{align}
    M_u = \langle H_u\rangle
    \begin{pmatrix}
        \alpha_1 (Y_\mathbf{3}^{(4)})_1, & \alpha_2^1 (Y_\mathbf{3}^{1,(8)})_1 + \alpha_2^2 (Y_\mathbf{3}^{2,(8)})_1, & \alpha_3^1 (Y_{\hat{\mathbf{3}}}^{1,(7)})_1 + \alpha_3^2 (Y_{\hat{\mathbf{3}}}^{2,(7)})_1\\
        \alpha_1 (Y_\mathbf{3}^{(4)})_3, & \alpha_2^1 (Y_\mathbf{3}^{1,(8)})_3 + \alpha_2^2 (Y_\mathbf{3}^{2,(8)})_3, & \alpha_3^1 (Y_{\hat{\mathbf{3}}}^{1,(7)})_3 + \alpha_3^2 (Y_{\hat{\mathbf{3}}}^{2,(7)})_3\\
        \alpha_1 (Y_\mathbf{3}^{(4)})_2, & \alpha_2^1 (Y_\mathbf{3}^{1,(8)})_2 + \alpha_2^2 (Y_\mathbf{3}^{2,(8)})_2, & \alpha_3^1 (Y_{\hat{\mathbf{3}}}^{1,(7)})_2 + \alpha_3^2 (Y_{\hat{\mathbf{3}}}^{2,(7)})_2\\
    \end{pmatrix}.
    \label{eq:mass_matrix_up}
\end{align}
In the same way, mass terms of the down-type quarks are derived as follows:
\begin{align}
    &\langle H_d\rangle \left\{ \beta_1 \left(Q_1 (Y_\mathbf{3}^{(4)})_1 +Q_2 (Y_\mathbf{3}^{(4)})_3 + Q_3 (Y_\mathbf{3}^{(4)})_2\right) d_1^c    \right\} 
   \nonumber\\
    &+\langle H_d\rangle \left\{ \beta_2 \left(Q_1 (Y_\mathbf{3}^{(6)})_1 +Q_2 (Y_\mathbf{3}^{(6)})_3 + Q_3 (Y_\mathbf{3}^{(6)})_2\right) d_2^c    \right\} 
   \nonumber\\
   &+\langle H_d\rangle \left\{ \sum_{a=1}^{2} \beta_3^a \left(Q_1 (Y_\mathbf{3}^{a,(8)})_1 + Q_2 (Y_\mathbf{3}^{a,(8)})_3 + Q_3 (Y_\mathbf{3}^{a,(8)})_2 \right) \right\}d_3^c,
\end{align}
i.e.,
\begin{align}
    M_d = \langle H_d\rangle
    \begin{pmatrix}
        \beta_1 (Y_\mathbf{3}^{(4)})_1, & \beta_2 (Y_\mathbf{3}^{(6)})_1, & \beta_3^1 (Y_\mathbf{3}^{1,(8)})_1 + \beta_3^2 (Y_\mathbf{3}^{2,(8)})_1\\
        \beta_1 (Y_\mathbf{3}^{(4)})_3, & \beta_2 (Y_\mathbf{3}^{(6)})_3, & \beta_3^1 (Y_\mathbf{3}^{1,(8)})_3 + \beta_3^2 (Y_\mathbf{3}^{2,(8)})_3\\
        \beta_1 (Y_\mathbf{3}^{(4)})_2, & \beta_2 (Y_\mathbf{3}^{(6)})_2, & \beta_3^1 (Y_\mathbf{3}^{1,(8)})_2 + \beta_3^2 (Y_\mathbf{3}^{2,(8)})_2\\
    \end{pmatrix}.
    \label{eq:mass_matrix_down}
\end{align}

\medskip

The K\"{a}hler potential of the chiral superfield $\phi_i$ is written as follows:
\begin{align}
K\supset \sum_i \frac{\phi_i^\dagger \phi_i}{\left( -i\tau+i\bar{\tau} \right)^{k_{\phi_i}}},
\end{align}
where $-k_{\phi_i}$ is a weight of the superfield. 
Thus, each component of the mass matrices is modified by the canonical normalization as 
\begin{align}
\left(M_\phi\right)_{ij} \to \left(\sqrt{2\Im\,[\tau]}\right)^{k_Q+k_{\phi_j}} \left(M_\phi\right)_{ij}, 
\end{align}
with $i,j = 1,2,3$ and $\phi=u,d$.  

\medskip

The mass matrices are diagonalized by unitary matrices as follows:
\begin{align}
    M_u &= U_{L}^{\dagger}\mathrm{diag}(m_u, m_c, m_t) U_{R},\\
    M_d &= V_{L}^{\dagger}\mathrm{diag}(m_d, m_s, m_b) V_{R}.
\end{align}
Moreover, the flavor mixing is defined as the difference between mass eigenstates and flavor eigenstates:
\begin{align} 
    U_{\mathrm{CKM}} &= U_{L} V_{L}^{\dagger} \nonumber \\
    &= \begin{pmatrix}
    c_{12} c_{13} & s_{12} c_{13} & s_{13} e^{-i\delta_{\rm{CP}}} \\ 
    -s_{12} c_{23} - c_{12} s_{23} s_{13} e^{i\delta_{\rm{CP}}} & c_{12} c_{23} - s_{12} s_{23} s_{13} e^{i\delta_{\rm{CP}}} & s_{23} c_{13} \\
    s_{12} s_{23} - c_{12} c_{23} s_{13} e^{i\delta_{\rm{CP}}} & -c_{12} s_{23} - s_{12} c_{23} s_{13} e^{i\delta_{\rm{CP}}} & c_{23} c_{13} 
    \end{pmatrix},
\end{align}
with $c_{ij} = \cos{\theta_{ij}}$ and $s_{ij} = \sin{\theta_{ij}}$.
Moreover, the Jarlskog invariant is defined as follows:
\begin{align}
    J = \Im\,\left[U_{11}U_{22}U_{12}^* U_{21}^*\right] = s_{23}c_{23}s_{12}c_{12}s_{13}c_{13}^2 \sin{\delta_{\mathrm{CP}}},
\end{align}
which is a measure of CP violation.

\medskip

At $\tan \beta = 10$ and the SUSY breaking scale $M_{\mathrm{SUSY}} = 10\,\TeV$, we show values of masses and mixings in Table \ref{tab:central_values}, which is calculated from Ref.~\cite{Antusch:2013jca}.
In addition, from mixing angles in that data, absolute values of the CKM matrix is derived as follows:
\begin{align}
    |U_{\mathrm{CKM}}| =
    \left(\begin{array}{ccc}
     0.9743 \pm 0.0002 & 0.2254 \pm 0.0007 & 0.0035 \pm 0.0001 \\
     0.2253 \pm 0.0007 & 0.9735 \pm 0.0002 & 0.0401 \pm 0.0006 \\
     0.0085 \pm 0.0002 & 0.0394 \pm 0.0006 & 0.9992 \pm 0.0000 \\
    \end{array}\right) .\label{eq:CKM_tanbeta10}
\end{align}

\medskip

From an analytical point of view, $\theta(\tau) \sim 1$ and $\epsilon(\tau) \sim 2q^{1/4} \ll 1$ when $2\pi \Im\,[\tau] \gg 1$.
This makes the hierarchical structure of mass matrices Eq.~\eqref{eq:mass_matrix_up} and Eq.~\eqref{eq:mass_matrix_down}, so it is easy to realize semi-realistic flavor structure.
It is non-trivial how small $\Im\,[\tau]$ is allowed to reproduce the flavor structure, and $\Im\,[\tau]\sim 2.8$ was found as one of the optimal values in Ref.~\cite{Abe:2023qmr}.

\begin{table}[H]
\centering
\begin{tabular}{|c|c|}
    \hline
     & $\mu_{i}\pm1\sigma$ \\
    \hline\hline
    $\left(m_u/m_t\right)/10^{-6}$   & $ 5.4412 \pm 1.7132 $ \\
    $\left(m_c/m_t\right)/10^{-3}$   & $ 2.8213 \pm 0.1195 $ \\
    $m_t/\GeV$  & $ 87.4555 $ \\
    \hline
    $\left(m_d/m_b\right)/10^{-4}$   & $ 9.2159 \pm 1.2382 $ \\
    $\left(m_s/m_b\right)/10^{-2}$   & $ 1.8241 \pm 0.1005 $ \\
    $m_b/\GeV$  & $ 0.9682 $ \\
    \hline
    $s_{12}^q/10^{-1}$ & $ 2.2736 \pm 0.0073 $ \\
    $s_{23}^q/10^{-2}$ & $ 4.015 \pm 0.064 $ \\
    $s_{13}^q/10^{-3}$ & $ 3.49 \pm 0.13 $ \\
    $\delta_{\text{CKM}}/\pi$ & $ 0.3845 \pm 0.0173 $ \\
    $J/10^{-5}$ & $ 2.87 \pm 0.13 $ \\
    \hline
\end{tabular}
\caption{The central values and $1\sigma$ ranges for the quark sector with $\tan \beta = 10$ and $M_{\mathrm{SUSY}} = 10\,\TeV$ based on Ref.~\cite{Antusch:2013jca}.} 
\label{tab:central_values}
\end{table}

\subsection{Conditional diffusion model}
\label{sec:diffusion_model}

In this section, we present the detailed design of the conditional diffusion model adopted for the $S_4^\prime$ modular flavor model as a case study.
The diffusion process, the inverse process, and the transfer learning will be described in this order.

\subsubsection*{Diffusion process}
\label{sec:diffusion_process}

We adopt $\tan \beta = 10.0,\,\alpha_3^1=0.001$.
In preparing the initial data, we deal with the following values:\footnote{In our analysis, the modulus field $\tau$ (symmetry breaking field) is regarded as a parameter, but the VEV of $\tau$ will be determined by its stabilization mechanism proposed in e.g., \cite{Ishiguro:2020tmo,Novichkov:2022wvg,Ishiguro:2022pde}.}
\begin{align}
    G &= \left\{ \frac{\alpha_1}{10\cdot\alpha_3^1},\,\frac{\alpha_2^1}{10\cdot\alpha_3^1},\,\frac{\alpha_2^2}{10\cdot\alpha_3^1},\,\frac{\alpha_3^2}{10\cdot\alpha_3^1},\,
    \frac{\beta_1}{10\cdot\alpha_3^1},\,\frac{\beta_2}{10\cdot\alpha_3^1},\,\frac{\beta_3^1}{10\cdot\alpha_3^1},\,\frac{\beta_3^2}{10\cdot\alpha_3^1},\,
    \Re\,[\tau],\,\Im\,[\tau] \right\}, \\
    L &= \left\{ \log_{10}\frac{m_u}{m_t},\,\log_{10}\frac{m_c}{m_t},\,0.0,\,\log_{10}\frac{m_d}{m_b},\,\log_{10}\frac{m_s}{m_b},\,0.0,\,|(U_{\rm CKM})_{ij}|,\,\mathrm{sign}[J]\times\log_{10}|J| \right\},
\end{align}
with $i,j=1,2,3$.
For $L$, we introduce trivial labels to represent $\log_{10}(m_t / m_t)$ and $\log_{10}(m_b / m_b)$. 
These dummy labels allow the dimensions of $L$ to be matched to a number of independent physical quantities, which is beneficial for developing a generic architecture of the neural network. 
However, it remains uncertain whether these dummy labels can allow the small $\chi^2$. 
The necessity of these dummy labels will be reported elsewhere. 
In addition, we avoid exponential differences in the input values to the neural network by applying the logarithm to the mass ratio and the Jarlskog invariant.

\medskip

Now, $G$ has 10 components and $L$ has 16 components.
Each element of $G$ was generated as uniform random numbers satisfying the following ranges:
\begin{align}
    0.01 &\leq \left\{\left|\frac{\alpha_1}{10\cdot\alpha_3^1}\right|,\,\left|\frac{\alpha_2^1}{10\cdot\alpha_3^1}\right|,\,\left|\frac{\alpha_2^2}{10\cdot\alpha_3^1}\right|,\,\left|\frac{\alpha_3^2}{10\cdot\alpha_3^1}\right|\right\} \leq 1.0, \label{eq:alpha_range} \\
    0.01 &\leq \left\{\left|\frac{\beta_1}{10\cdot\alpha_3^1}\right|,\,\left|\frac{\beta_2}{10\cdot\alpha_3^1}\right|,\,\left|\frac{\beta_3^1}{10\cdot\alpha_3^1}\right|,\,\left|\frac{\beta_3^2}{10\cdot\alpha_3^1}\right|\right\} \leq 1.0, \label{eq:beta_range} \\
    -\frac{1}{2} &\leq \Re\,[\tau] \leq \frac{1}{2}, \quad
    \frac{\sqrt{3}}{2} \leq \Im\,[\tau] \leq 5.0.
    \label{eq:tau_range}
\end{align}
Eq.~\eqref{eq:alpha_range} and Eq.~\eqref{eq:beta_range} mean that the ratios of coefficients $r$ satisfy $0.1 \leq |r| \leq 10$.
In addition, data processing techniques such as normalization and standardization are important for achieving efficient learning of neural networks. 
In this respect, the ratio $r$ in the component of $G$ is divided by ten to realize the proper normalization $r\sim \mathcal{O}(1)$.

\medskip

Then, $L$ is computed from a randomly generated $G$ based on Eq.~\eqref{eq:mass_matrix_up} and Eq.~\eqref{eq:mass_matrix_down}, and a neural network is trained using pairs of the input and the label $x_{0} = \left(G,L\right)$.
In actual learning process, we prepare $10^5$ pairs of $\left(G,L\right)$ as initial data.
Of these data, 90\% are used as training data and 10\% as validation data.
When the integer $t$ is randomly selected from $[1,T]$, the noisy data $x_{t}$ is determined according to Eq.~\eqref{eq:data_sequence}. 
When an input layer of the neural network receives $x_{t}$ as input, it performs nonlinear transformations according to Eq.~\eqref{eq:activation}. 

\medskip

The detailed architecture of our neural network is presented in Table \ref{tab:network}. 
There are 646,836 parameters that are adjusted during training. 
The activation function $h$ in Eq.~\eqref{eq:activation} is selected as the SELU function for hidden layers and an identity function for an output layer. 
The batch size is set to 64, and the parameter updates are performed $10^5$ times.
In addition, we employ the ADAM optimizer in PyTorch to update the weight $w$ and bias $b$ described in Eq.~\eqref{eq:activation}. 
The learning rate is automatically adjusted using the scheduler ``OneCycleLR'' provided by PyTorch, and we adopted $r_{\mathrm{max}}=0.001$ as the maximum rate. 

\begin{table}[H]
    \centering
    \begin{tabular}{|c|c|c||c|c|c|}
    \hline
    Layer & Dimension & Activation & Layer & Dimension & Activation \\
    \hline \hline
    Input & $\mathbb{R}^{10}_{x} + \mathbb{R}^{17}_{L,t}$ & - & Hidden 6 & $\mathbb{R}^{512} + \mathbb{R}^{17}_{L,t}$ & \text{SELU} \\
    Hidden 1 & $\mathbb{R}^{32} + \mathbb{R}^{17}_{L,t}$ & \text{SELU} & Hidden 7 & $\mathbb{R}^{256} + \mathbb{R}^{17}_{L,t}$ & \text{SELU} \\
    Hidden 2 & $\mathbb{R}^{64} + \mathbb{R}^{17}_{L,t}$ & \text{SELU} & Hidden 8 & $\mathbb{R}^{128} + \mathbb{R}^{17}_{L,t}$ & \text{SELU} \\
    Hidden 3 & $\mathbb{R}^{128} + \mathbb{R}^{17}_{L,t}$ & \text{SELU} & Hidden 9 & $\mathbb{R}^{64} + \mathbb{R}^{17}_{L,t}$ & \text{SELU} \\
    Hidden 4 & $\mathbb{R}^{256} + \mathbb{R}^{17}_{L,t}$ & \text{SELU} & Hidden 10 & $\mathbb{R}^{32} + \mathbb{R}^{17}_{L,t}$ & \text{SELU} \\
    Hidden 5 & $\mathbb{R}^{512} + \mathbb{R}^{17}_{L,t}$ & \text{SELU} & Output & $\mathbb{R}^{10}$ & \text{Identity} \\
    \hline
    \end{tabular}
    \caption{In the neural network, the inputs are the noised data $x_{t}$ (10 dimensions), the label $L$ (16 dimensions), and the time $t$. The activation functions are the SELU function for hidden layers and the identity function for the output layer. $L$ is treated as $\varnothing$ in probability of 10\%.}
    \label{tab:network}    
\end{table}

\subsubsection*{Reverse process}
\label{sec:reverse_process}

In the reverse process, the generation is performed with labels that reflect real observables. 
In particular, $L$ is designated as follows based on Table \ref{tab:central_values} and Eq.~\eqref{eq:CKM_tanbeta10}:
\begin{align}
\begin{split}
    &\log_{10}\frac{m_u}{m_t} = \log_{10} \left(5.4412 \times 10^{-6}\right) = -5.2643, \\
    &\log_{10}\frac{m_c}{m_t} = \log_{10} \left(2.8213 \times 10^{-3}\right) = -2.5496, \\
    &\log_{10}\frac{m_d}{m_b} = \log_{10} \left(9.2159 \times 10^{-4}\right) = -3.0355, \\
    &\log_{10}\frac{m_s}{m_b} = \log_{10} \left(1.8241 \times 10^{-2}\right) = -1.7390, \\
    &|(U_{\rm CKM})_{ij}| =
    \left(\begin{array}{ccc}
     0.9743 & 0.2254 & 0.0035 \\
     0.2253 & 0.9735 & 0.0401 \\
     0.0085 & 0.0394 & 0.9992 \\
    \end{array}\right), \\
    &\mathrm{sign}[J]\times\log_{10}|J| = +\log_{10} \left(2.87 \times 10^{-5}\right) = -4.54.
\end{split}
\end{align}
Then, $L$ is recalculated using the generated $G$, and the data that falls within a specified error range is extracted.
As a result, the data $G$ corresponding to the experimental values has been derived solely from the experimental label.

\subsubsection*{Transfer learning}
\label{sec:transfer}

After the diffusion model generates the data $G$, the physical values $P_{\ell}$ corresponding to $G$ can be calculated. 
The accuracy of $P_{\ell}$ in comparison to the target experimental label $L$ is quantitatively evaluated by the $\chi^{2}$ value defined as follows:
\begin{align}
    \chi^{2} = \sum_{\ell=1}^{n} \left( \frac{P_{\ell}-\mu_{\ell}}{\sigma_{\ell}} \right)^{2},
\end{align}
where $P_{\ell}, \mu_{\ell}, \sigma_{\ell}$ respectively denote prediction for physical observable, central value and $1\sigma$ deviations.
In this work, the $\chi^{2}$ function is calculated for following 8 observables:
\begin{align}
    \left\{ \frac{m_u}{m_t},\,\frac{m_c}{m_t},\,\frac{m_d}{m_b},\,\frac{m_s}{m_b},\,
    \theta_{12}^q,\,\theta_{23}^q,\,\theta_{13}^q,\,\left|\frac{\delta_{\mathrm{CKM}}}{\pi}\right| \right\}, \label{eq:def_chisq}
\end{align}
so the CP violation is considered in $\chi^{2}$.

\medskip

Now, we refer to the first neural network as a \textit{pre-network}. 
The data generated by the pre-network are collected for transfer learning, and the pre-network is retrained based on this new data. 
All parameters of the pre-network are updated, which is why the second training phase is referred to as fine-tuning. 
In our training process, the hyperparameters during the second phase are the same as those used in the first phase. 
The second network constructed through fine-tuning is referred to as a \textit{tuned-network} in the following discussion.

\newpage
\section{Results}
\label{sec:result}

Our calculations are performed using Google Colaboratory with a CPU (not a GPU), so this method accessible to everyone. 
The diffusion process takes approximately 1 hour, and the reverse process requires approximately 4.6 hours per generating $10^5$ data in a single session. 
In this section, we denote the chi-squared value for the data generated by the diffusion model itself as $\chi_{\mathrm{gen}}^2$.
First, we generate data using a pre-network. 
The total number of data generated by the network is $4\times10^6$, of which 103,663 satisfy the condition $\chi_{\mathrm{gen}}^2<8.0\times10^4$. 
Therefore, in the diffusion model \textbf{without} fine-tuning, the ratio of data satisfying condition $\chi_{\mathrm{gen}}^2<8.0\times10^4$ is 2.59\%. 
Second, we perform fine-tuning on the 103,663 data generated by the pre-network that satisfy condition $\chi_{\mathrm{gen}}^2<8.0\times10^4$. 
When we generate a total of $9\times10^6$ data using the tuned-network, there are $\{928, 110, 17\}$ solutions that satisfy the condition $\chi_{\mathrm{gen}}^2<\{600,400,200\}$. 
Finally, we extract $\{611, 58, 11\}$ solutions such that the CP phase is positive.
Here, in the diffusion model \textbf{with} fine-tuning, the ratio of data satisfying condition $\chi_{\mathrm{gen}}^2<8.0\times10^4$ is 5.95\%.

\medskip

Based on these results, fine-tuning achieves a smaller $\chi^2_{\rm gen}$ compared to the case without fine-tuning.
Furthermore, since the architectures of both the pre-network and the tuned-network are identical, small $\chi_{\mathrm{gen}}^2$ can be achieved simply by repeating the learning process. 
Consequently, this method of improvement can be applied irrespective of the specific details of the flavor model. 
It is also anticipated that fine-tuning can be repeated to achieve the desired level of accuracy.

\medskip

\begin{figure}[H]
    \centering
    \includegraphics[height=60mm]{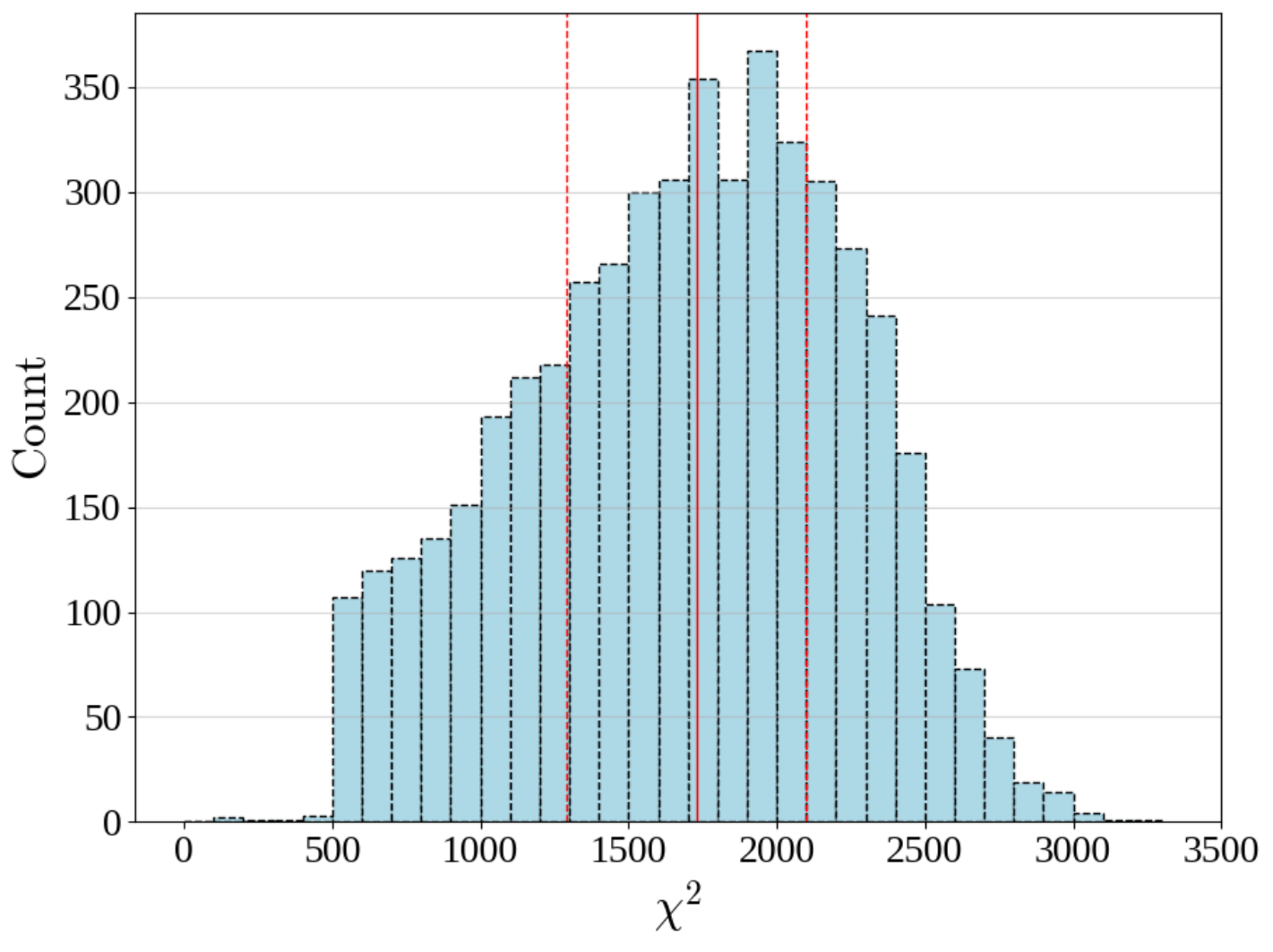}
    \caption{The distribution of the $\chi^2$ values obtained from the MCMC-optimized test is shown. The red line indicates the median value (1,732), while the dotted lines represent the first and third quartiles (1,290 and 2,101, respectively). Note that this calculation does not use the diffusion model.}
    \label{fig:mcmc_refinement}
\end{figure}

In the calculations described above using the diffusion model, a total of 600 hours of computation is performed to obtain 611 solutions with $\chi_{\mathrm{gen}}^2<600$. 
This result is remarkable in two respects. 
First, when generating $4\times10^6$ random initial values within the ranges specified by Eqs.~\eqref{eq:alpha_range}, \eqref{eq:beta_range}, and \eqref{eq:tau_range}, the minimum, first quartile, median, third quartile, and maximum $\chi^2$ values are respectively given by $\{ 4.00 \times 10^{3}, 9.89 \times 10^{4}, 4.92 \times 10^{5}, 1.98 \times 10^{7}, 2.31 \times 10^{11} \}$. 
Optimizing these initial values within the same computational time as the diffusion model requires processing at 0.54 seconds per data point, rendering this approach impractical. 
Second, to compare its performance with conventional methods, global optimization via MCMC is also attempted. 
In this process, 5,000 random initial values are prepared within the same parameter ranges, and 440 seconds are spent at each point, resulting in a total computational cost of 610 hours. 
The $\chi^2$ distribution from the optimization has a median value of 1,732 and a minimum value of 125, as shown in Fig.~\ref{fig:mcmc_refinement}. 
Among these, $\{114, 4, 2\}$ solutions satisfy the condition $\chi_{\mathrm{gen}}^2<\{600, 400, 200\}$, respectively. 
The significant difference in the number of solutions obtained within equivalent computation time demonstrates the advantage of global search using the diffusion model. 
While it is possible to improve the efficiency of MCMC exploration, doing so requires restricting the search space, selecting appropriate initial values, and adjusting exploration parameters. 
In contrast, diffusion model-based optimization eliminates the need for such tuning, thereby significantly reducing the effort required for exploration.

\medskip

\begin{figure}[H]
 
 \begin{minipage}{0.3\hsize}
  \begin{center}
  \includegraphics[height=45mm]{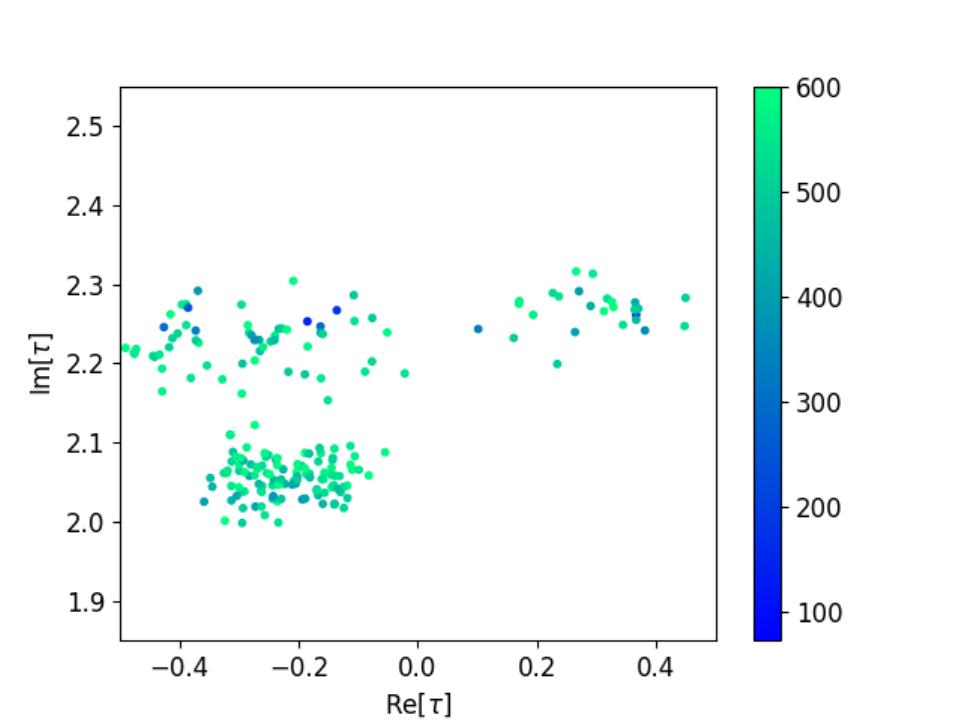}
  \end{center}
\end{minipage}
\begin{minipage}{0.3\hsize}
  \begin{center}
  \includegraphics[height=45mm]{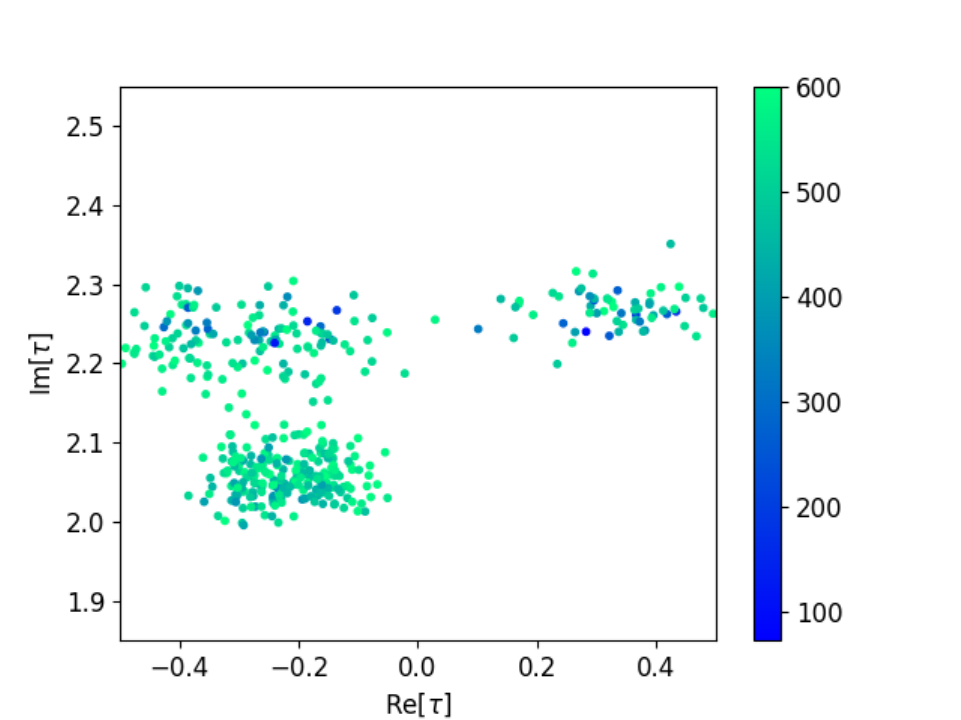}
  \end{center}
\end{minipage}
\begin{minipage}{0.3\hsize}
  \begin{center}
  \includegraphics[height=45mm]{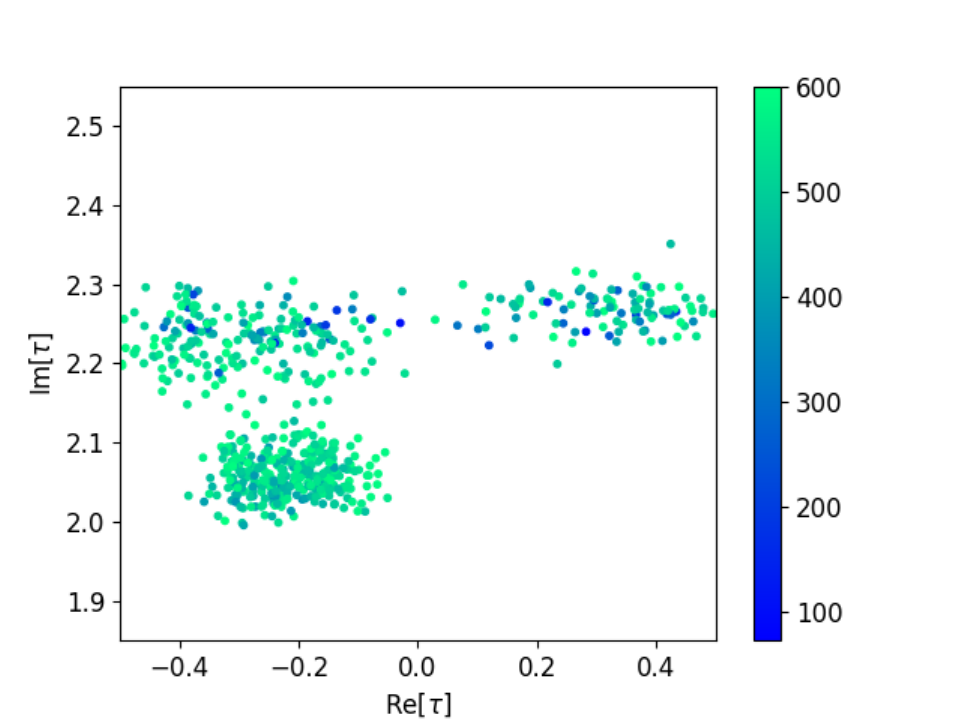}
  \end{center}
\end{minipage}
\\
 \begin{minipage}{0.3\hsize}
  \begin{center}
  \includegraphics[height=45mm]{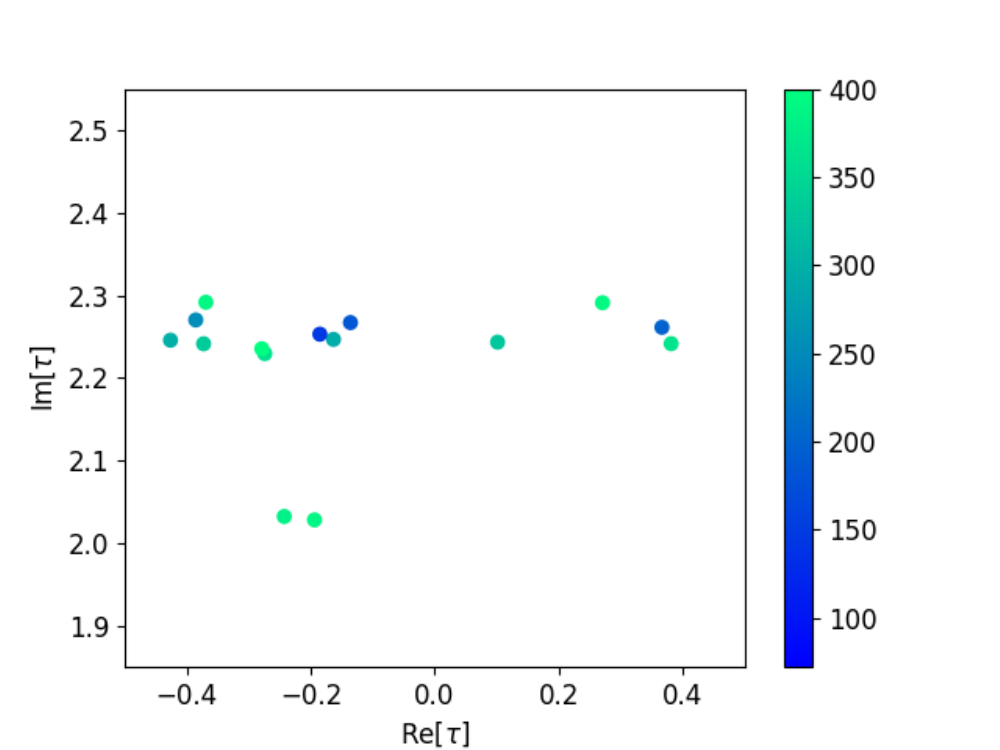}
  \end{center}
\end{minipage}
\begin{minipage}{0.3\hsize}
  \begin{center}
  \includegraphics[height=45mm]{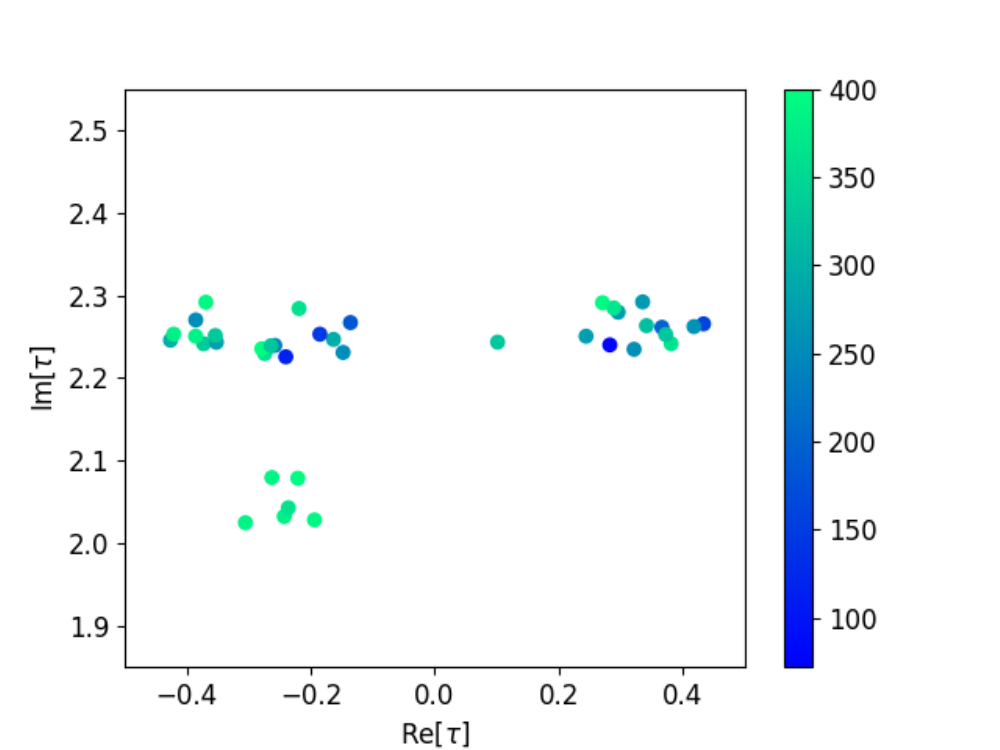}
  \end{center}
\end{minipage}
\begin{minipage}{0.3\hsize}
  \begin{center}
  \includegraphics[height=45mm]{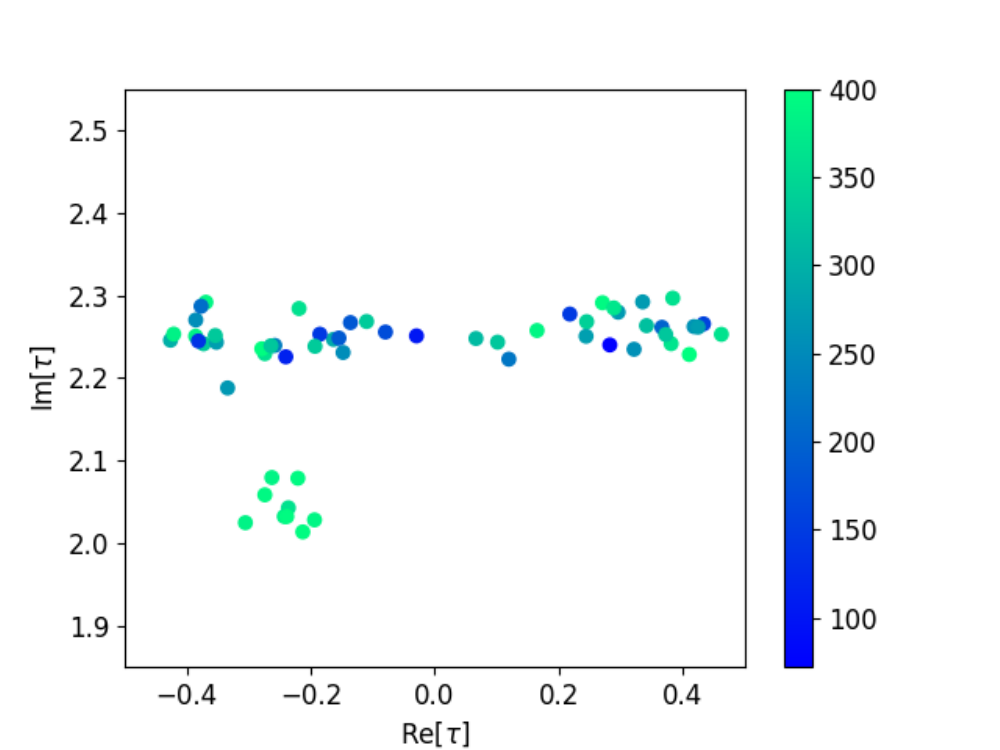}
  \end{center}
\end{minipage}
\\
\begin{minipage}{0.3\hsize}
  \begin{center}
  \includegraphics[height=45mm]{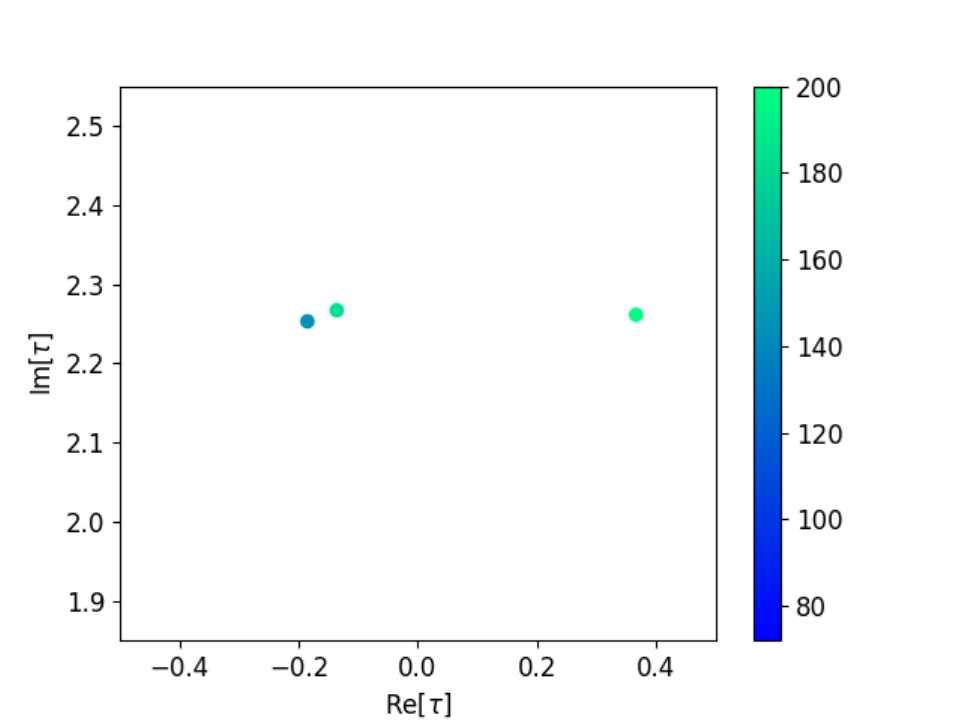}
  \end{center}
\end{minipage}
\begin{minipage}{0.3\hsize}
  \begin{center}
  \includegraphics[height=45mm]{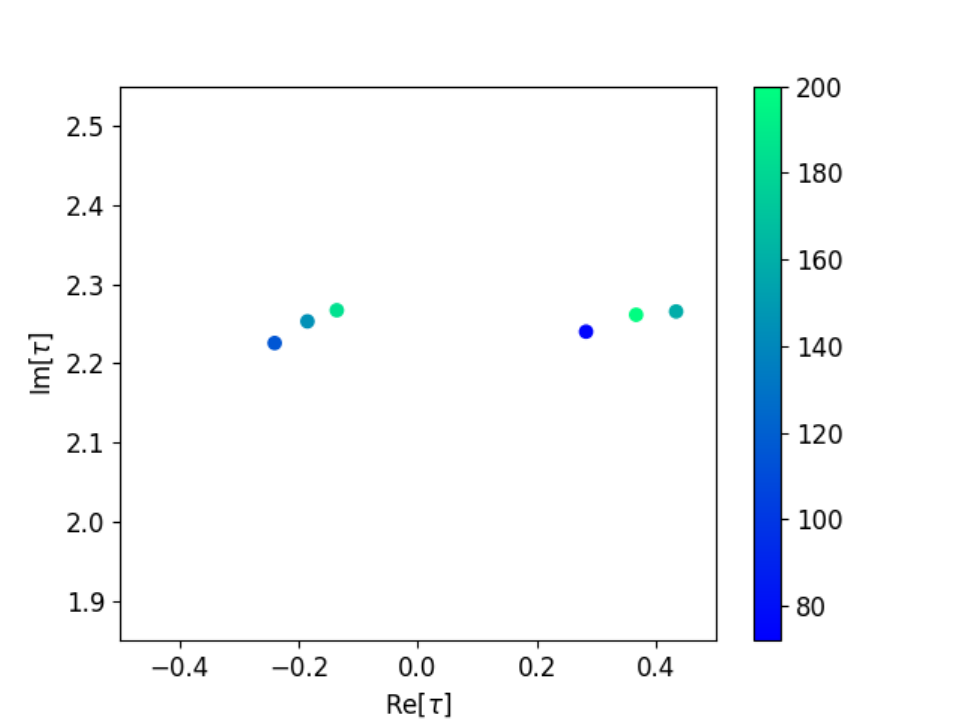}
  \end{center}
\end{minipage}
\begin{minipage}{0.3\hsize}
  \begin{center}
  \includegraphics[height=45mm]{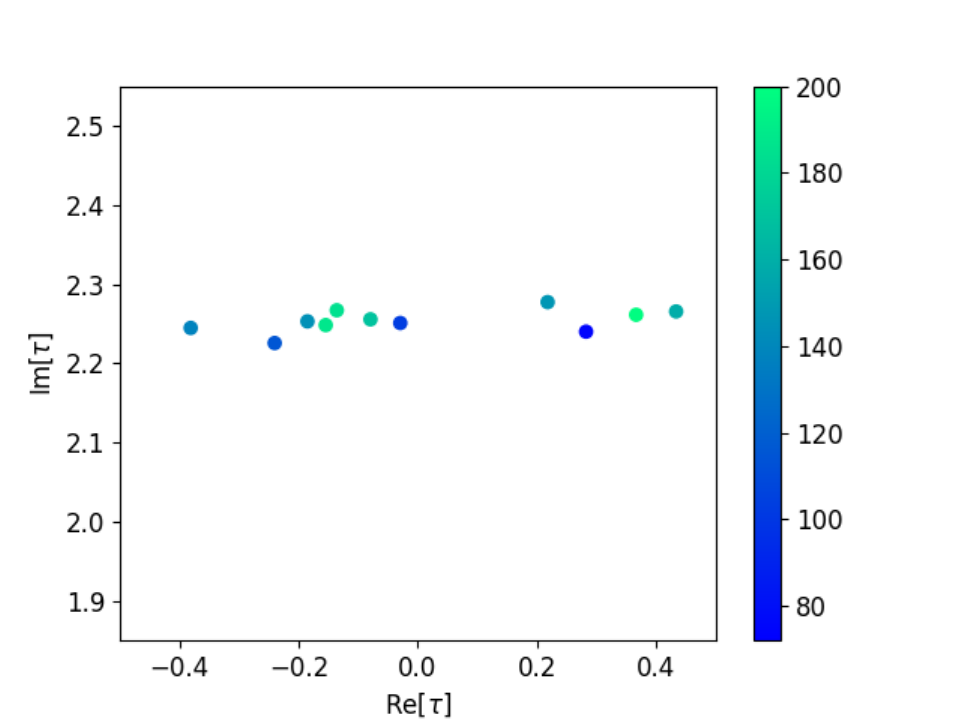}
  \end{center}
\end{minipage}

  \caption{In the upper row, distributions of modulus $\tau$ that satisfy $\chi_{\mathrm{gen}}^2<600$ among the data generated by the diffusion model are displayed, and there are 611 candidates. On the other hand, in the middle (lower) row, we show the distributions in the 58 (11) solutions that satisfy $\chi_{\mathrm{gen}}^2<400$ ($\chi_{\mathrm{gen}}^2<200$), which reproduce the experimental values with relatively small $\chi_{\mathrm{gen}}^2$. The left, middle, and right figures correspond to the points when the total number of generated data is $3\times10^6$, $6\times10^6$, and $9\times10^6$, respectively.}
\label{fig:res_progress}
\end{figure}

To show the progress of the generation process, Fig.~\ref{fig:res_progress} presents three graphs depicting the distribution of modulus $\tau$ as the amount of generated data increases. 
We observed a growing number of candidates that reproduce the flavor structure of the quarks with $\chi_{\mathrm{gen}}^2<600$. 
In fact, phenomenologically promising candidates of $\tau$ appear in various locations, particularly concentrated around $\Im\,[\tau] \sim 2.2$. 
The values of the data $G$ corresponding to the solution with the lowest $\chi^2$ value are as follows:
\begin{align}
\begin{split}
    &\frac{\alpha_1}{\alpha_3^1}=0.0386,\quad\frac{\alpha_2^1}{\alpha_3^1}=0.0479,\quad\frac{\alpha_2^2}{\alpha_3^1}=0.4705,\quad\frac{\alpha_3^2}{\alpha_3^1}=-1.1378, \\
    &\frac{\beta_1}{\alpha_3^1}=-4.3717,\quad\frac{\beta_2}{\alpha_3^1}=-0.2171,\quad\frac{\beta_3^1}{\alpha_3^1}=2.8057,\quad\frac{\beta_3^2}{\alpha_3^1}=-7.5297, \\
    &\Re\,[\tau]=0.2825, \quad
    \Im\,[\tau]=2.2400, \label{eq:best_solution}
\end{split}
\end{align}
with $\chi_{\mathrm{gen}}^2=74.4$.
A more accurate solution can also be obtained by optimizing around the solutions generated by the diffusion model. 
Here, the values shown in Eq.~\eqref{eq:best_solution} are optimized using MCMC method. 
When optimization is performed within a search range of $\pm10\%$ around each value in the equation, the following solution is obtained after only 10 minutes of computation using Google Colaboratory.
\begin{align}
\begin{split}
    &\frac{\alpha_1}{\alpha_3^1}=0.0393,\quad\frac{\alpha_2^1}{\alpha_3^1}=0.0519,\quad\frac{\alpha_2^2}{\alpha_3^1}=0.4426,\quad\frac{\alpha_3^2}{\alpha_3^1}=-1.2094, \\
    &\frac{\beta_1}{\alpha_3^1}=-4.6663,\quad\frac{\beta_2}{\alpha_3^1}=-0.2095,\quad\frac{\beta_3^1}{\alpha_3^1}=3.0438,\quad\frac{\beta_3^2}{\alpha_3^1}=-8.1829, \\
    &\Re\,[\tau]=0.2828, \quad
    \Im\,[\tau]=2.2434. \label{eq:optimized_solution}
\end{split}
\end{align}
Using the notation $\chi_{\mathrm{opt}}^2$ to denote the chi-squared value of the optimized data, i.e., $\chi^2$ obtained through the diffusion model with MCMC, these parameters lead to the following observables within $\chi_{\mathrm{opt}}^2=8.42$:
\begin{align}
\begin{split}
    \left( \frac{m_u}{m_t}\big/10^{-6},\,\frac{m_c}{m_t}\big/10^{-3} \right)
    &= \left( 0.704,\, 2.82 \right), \\
    \left( \frac{m_d}{m_b}\big/10^{-4},\,\frac{m_s}{m_b}\big/10^{-2} \right)
    &= \left( 9.31,\, 1.81 \right), \\
    \left( s_{12}^q/10^{-1},\,s_{23}^q/10^{-2},\,s_{13}^q/10^{-3},\,\delta_{\mathrm{CKM}}/\pi \right)
    &= \left( 2.27,\, 4.06,\, 3.5,\, 0.377 \right).
\end{split}
\end{align}
Thus, by using the solution obtained from the diffusion model as the initial value for optimization, human beings no longer need to narrow the search range in advance and can instead focus on examining the areas proposed by the machine.

\medskip

\begin{figure}[H]
 \begin{minipage}{0.5\hsize}
  \begin{center}
  \includegraphics[height=60mm]{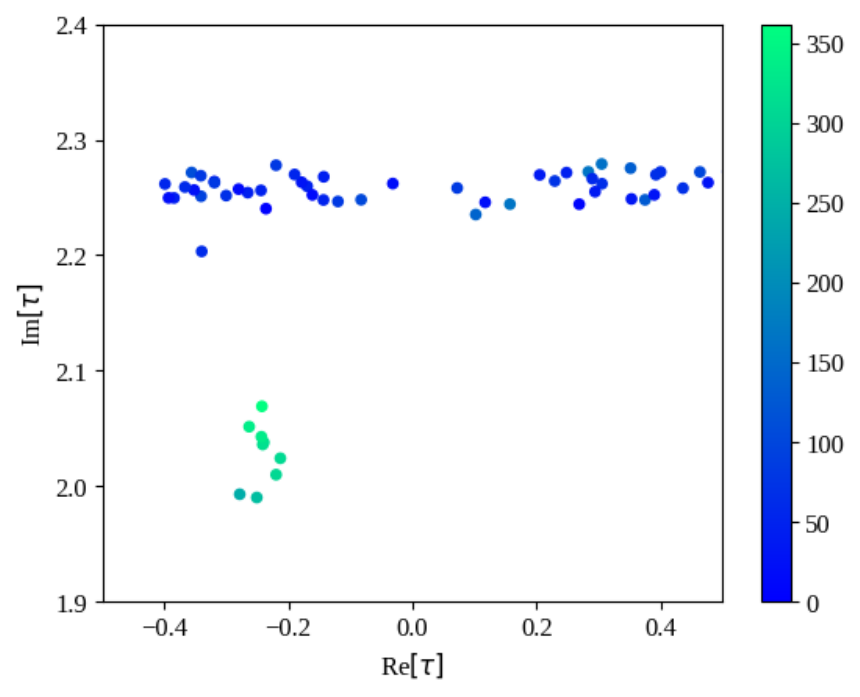}
  \end{center}
\end{minipage}
\begin{minipage}{0.5\hsize}
  \begin{center}
  \includegraphics[height=62mm]{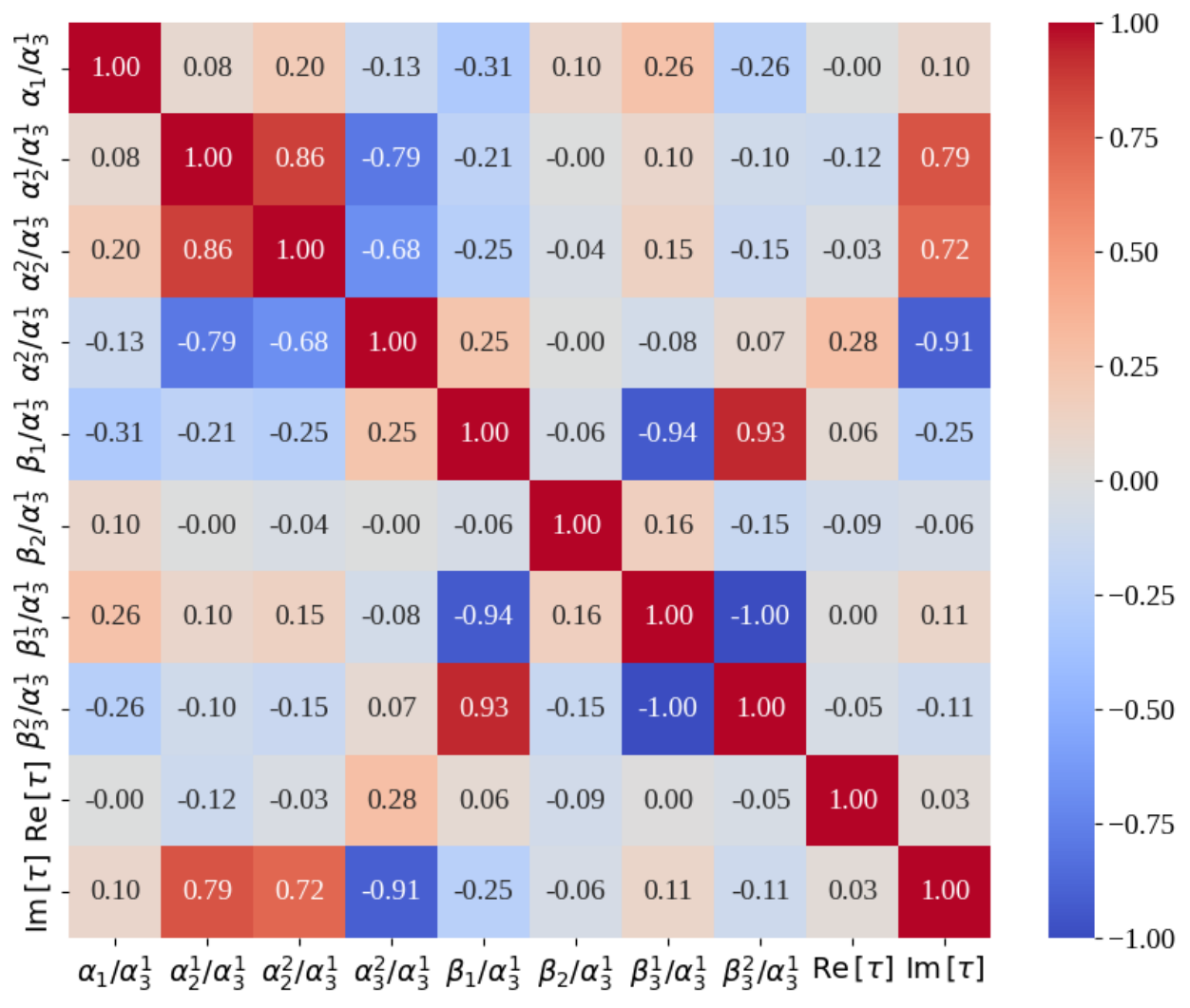}
  \end{center}
\end{minipage}
  \caption{The left graph shows the distributions of the modulus $\tau$ obtained by the diffusion model with MCMC. These positions, derived from 58 solutions with $\chi_{\mathrm{gen}}^2<400$, do not differ significantly from the original distributions shown in Fig.~\ref{fig:res_progress}. Other candidates with $\chi_{\mathrm{gen}}^2<600$ exhibit the same trend. The correlation of the parameters is illustrated in the figure on the right, showing the 139 points that satisfy the condition $\chi_{\mathrm{opt}}^2<50$ out of the 611 points. For example, $\Im\,[\tau]$ is strongly correlated with $\left\{ \alpha_2^1, \alpha_2^2, \alpha_3^2 \right\}$.}
\label{fig:optimized_distribution}
\end{figure}

Indeed, the following best fit with $\chi_{\mathrm{opt}}^2=1.00$ is also found after the optimization from the solution with $\chi_{\mathrm{gen}}^2=115$ generated by the diffusion model:
\begin{align}
\begin{split}
    &\frac{\alpha_1}{\alpha_3^1}=-0.4882,\quad\frac{\alpha_2^1}{\alpha_3^1}=0.0745,\quad\frac{\alpha_2^2}{\alpha_3^1}=0.4692,\quad\frac{\alpha_3^2}{\alpha_3^1}=-1.2691, \\
    &\frac{\beta_1}{\alpha_3^1}=-3.6368,\quad\frac{\beta_2}{\alpha_3^1}=-0.1516,\quad\frac{\beta_3^1}{\alpha_3^1}=2.3802,\quad\frac{\beta_3^2}{\alpha_3^1}=-6.2575, \\
    &\Re\,[\tau]=-0.2251, \quad
    \Im\,[\tau]=2.2423. \label{eq:best_fit}
\end{split}
\end{align}
These parameters lead to the following observables:
\begin{align}
\begin{split}
    \left( \frac{m_u}{m_t}\big/10^{-6},\,\frac{m_c}{m_t}\big/10^{-3} \right)
    &= \left( 6.24,\, 2.79 \right), \\
    \left( \frac{m_d}{m_b}\big/10^{-4},\,\frac{m_s}{m_b}\big/10^{-2} \right)
    &= \left( 8.66,\, 1.78 \right), \\
    \left( s_{12}^q/10^{-1},\,s_{23}^q/10^{-2},\,s_{13}^q/10^{-3},\,\delta_{\mathrm{CKM}}/\pi \right)
    &= \left( 2.28,\, 3.99,\, 3.5,\, 0.384 \right).
\end{split}
\end{align}
The distribution of optimized solutions and the correlations among the parameters are shown in Fig.~\ref{fig:optimized_distribution}. 
Regarding the $\tau$ plane, while the vertical axis focuses on the narrow region, high precision has been achieved with minimal deviation from the original distribution shown in Fig.~\ref{fig:res_progress}. 
In addition, although there are solutions near $\Im\,[\tau]=2.0$ in Fig.~\ref{fig:optimized_distribution}, these results are difficult to achieve with high precision even when using local optimization.
Examining the correlation coefficients for the model parameters reveals that among $\left\{ \alpha_1, \alpha_2^1, \alpha_2^2, \alpha_3^2 \right\}$, $\alpha_1$ exhibits relatively weak correlations with the other parameters. 
Similarly, $\beta_2$ also has weak correlations within $\left\{ \beta_1, \beta_2, \beta_3^1, \beta_3^2 \right\}$. 
Furthermore, it is evident that $\Im\,[\tau]$ strongly correlates with $\left\{ \alpha_2^1, \alpha_2^2, \alpha_3^2 \right\}$.

\newpage
\vspace*{\stretch{1}}

\vspace{\stretch{1}}
\newpage

\medskip

\begin{figure}[H]
 \begin{minipage}{0.5\hsize}
  \begin{center}
  \includegraphics[height=50mm]{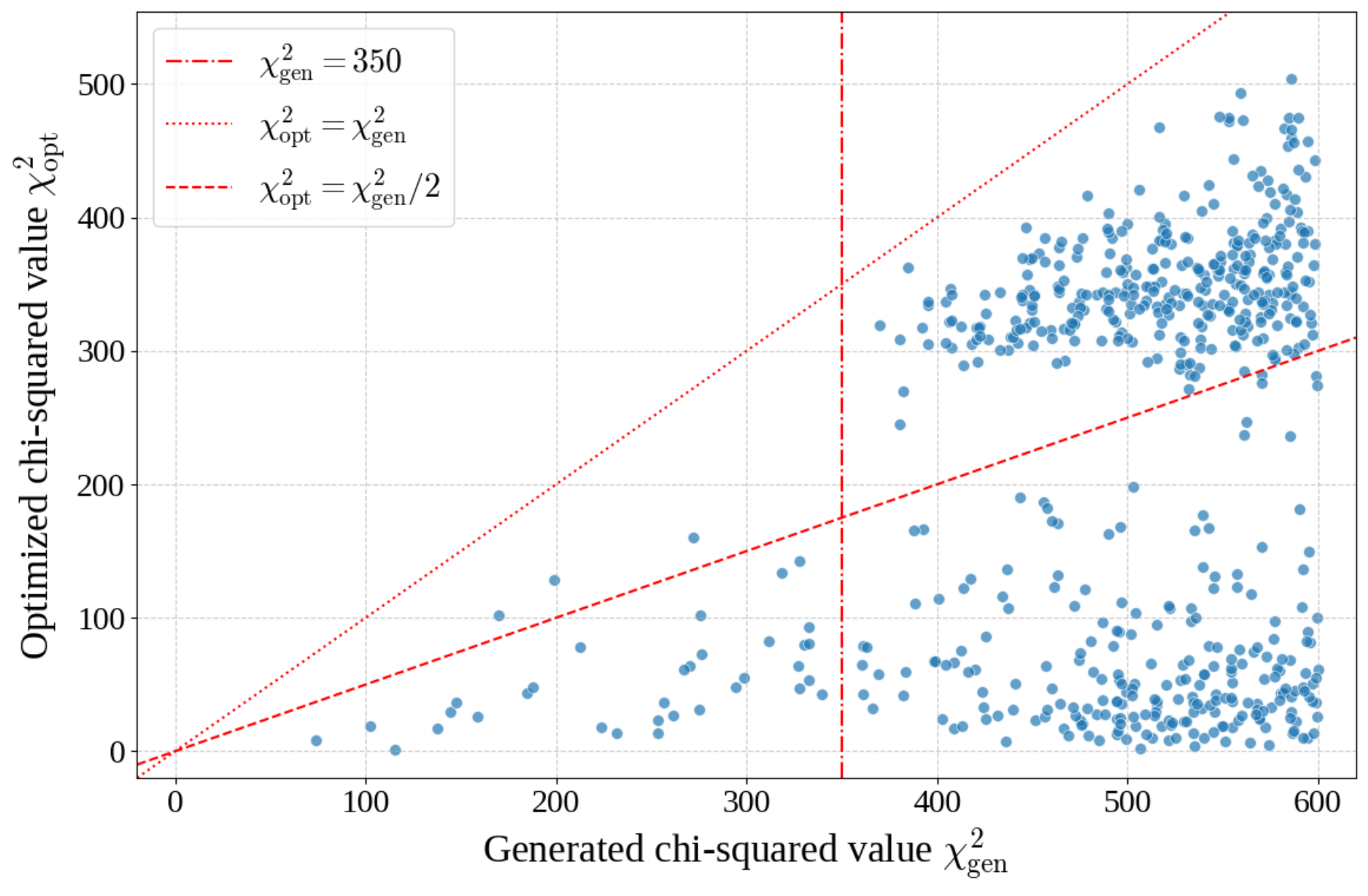}
  \end{center}
\end{minipage}
\begin{minipage}{0.5\hsize}
  \begin{center}
  \includegraphics[height=50mm]{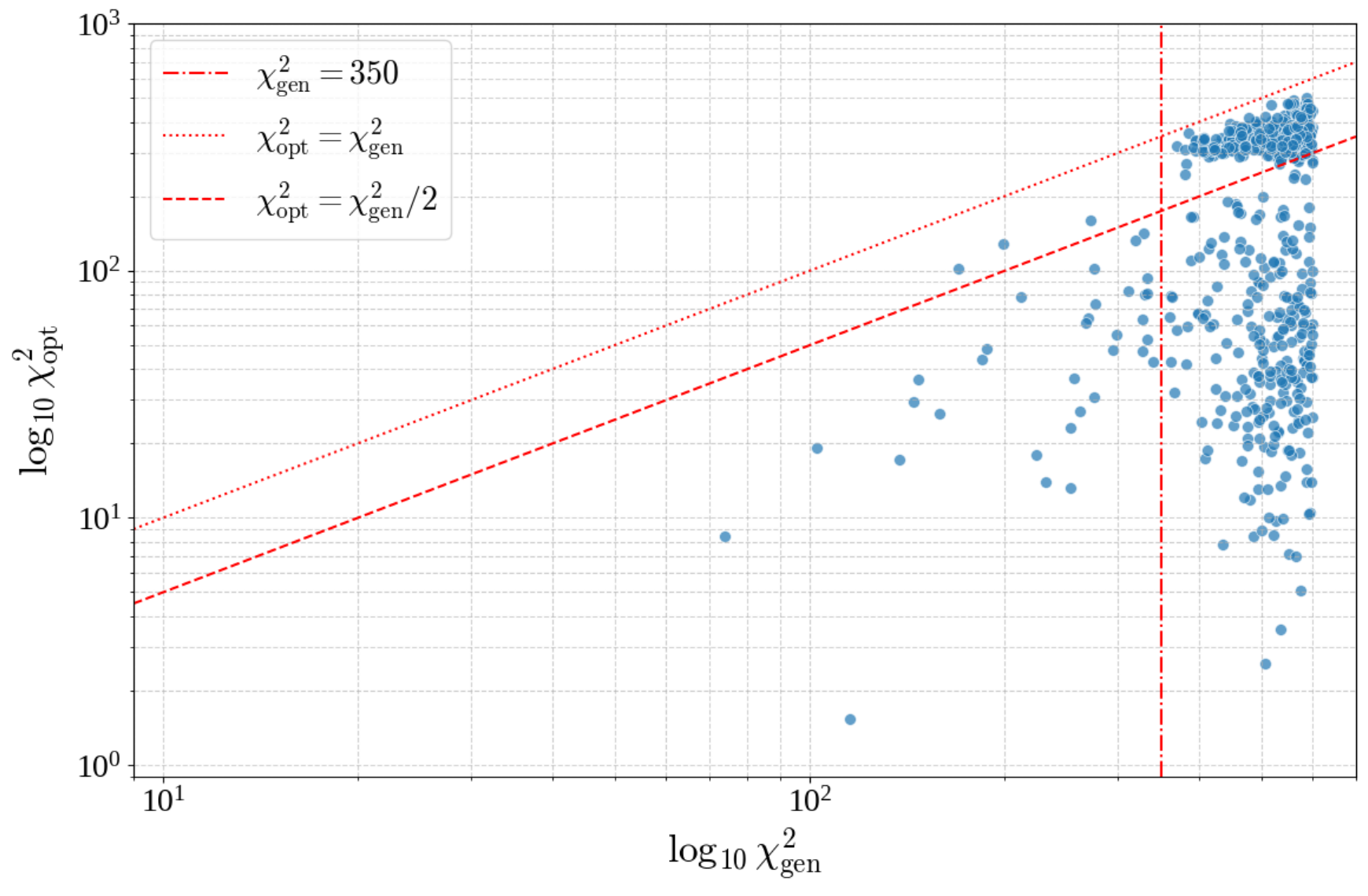}
  \end{center}
\end{minipage}
\caption{As shown in the left panel, a scatter plot of the 611 data points generated by the diffusion model is presented, with the original accuracy $\chi_{\mathrm{gen}}^2$ on the horizontal axis and the optimized accuracy $\chi_{\mathrm{opt}}^2$ on the vertical axis. The right panel shows a log-log version of the plot. The dash-dot line indicates the position where $\chi_{\mathrm{gen}}^2=350$; the behavior of the optimization changes approximately at this boundary. In particular, most points with $\chi_{\mathrm{gen}}^2 \lesssim 350$ satisfy $\chi_{\mathrm{opt}}^2 \leq \chi_{\mathrm{gen}}^2/2$.}
\label{fig:threshold}
\end{figure}

Fig.~\ref{fig:threshold} illustrates the relationship between the initial chi-squared value ($\chi^2_{\text{gen}}$) and the optimized value ($\chi^2_{\text{opt}}$) for each data point generated by the diffusion model. As shown in the figure, $\chi^2_{\text{gen}}$ systematically decreases after optimization, indicating that the diffusion model not only produces approximate solutions but also generates ``valid initial values'' capable of converging to realistic solutions. Notably, in the region where $\chi^2_{\text{gen}} \lesssim 350$, the majority of generated points exhibit significant improvement in accuracy, with $\chi^2_{\text{opt}} \leq \chi^2_{\text{gen}} / 2$ following local optimization. This result demonstrates that by using generated points satisfying a specific criterion (in this case, $\chi^2_{\text{gen}} \lesssim 350$) as starting points for the MCMC search, we can efficiently and systematically converge to high-precision solutions.

\medskip

As mentioned above, the generated results indicate that the transfer learning allows us to obtain the small $\chi^2_{\rm gen}$ compared to the case without the transfer learning. 
In order to reproduce the experimental values with even greater precision, it may be beneficial not only to conduct transfer learning multiple times but also to combine conventional methods, such as the Monte-Carlo method, with the parameters proposed by the diffusion model. 
Although we adopt the latter approach in this paper, the superiority of either approach remains to be investigated, and a comparison of their effectiveness taking into account computational resources should be reserved for future research.

\medskip

As described in Sec.~\ref{sec:S4prime_quark}, the flavor structure in the $S_4^\prime$ modular flavor model strongly depends on $\Im\,[\tau]$. 
When $\Im\,[\tau]$ is large, it is easy to reproduce the semi-realistic flavor structure. 
On the other hand, in regions where $\Im\,[\tau]$ is small, estimating the appropriate $\Im\,[\tau]$ becomes challenging, as even a slight variation can lead to a significantly different flavor structure. 
As a result, analytic evaluation with small $\Im\,[\tau]$ is a difficult issue. 
In addition, appropriate search ranges and increments of free parameters must be set for numerical optimization with conventional methods. 
These preparatory work are highly dependent on each researcher and require a trial-and-error process to achieve good settings.
On the other hand, the diffusion model that we developed does not impose strict restrictions on the parameter regions to be explored and generates a variety of candidates across a broad search range. 
Because of this property, there is no need for human beings to adjust the search region of parameters. 
Therefore, it is concluded that diffusion models with MCMC can automatically extract promising solutions from a wide search space using a procedure that is independent of the model's specifics.

\medskip

We now discuss the physical implications of the top 10 results generated by the diffusion model. 
Table \ref{tab:solutions} summarizes the values of $\tau$ and the Jarlskog invariant for each result. 
In particular, the median value of the Jarlskog invariant is $2.83 \times 10^{-5}$, which is comparable in magnitude to the experimental value $2.87 \times 10^{-5}$. 
Various previous studies addressing the $S_4^\prime$ modular flavor models often introduce the coefficients $\{\alpha,\beta\}$ as complex numbers to reproduce a Jarlskog invariant of reasonable size. 
In contrast, this study treats the parameters $\{\alpha,\beta\}$ as real numbers, and the Jarlskog invariant is reproduced from only $\Re\,[\tau]$. 
Thus, the spontaneous CP violation can be realized in the $S_4^\prime$ modular flavor model. 

\medskip

Regarding spontaneous CP violation, the ratios among coefficients are also relevant. 
For example, in Ref.~\cite{Abe:2023qmr}, the coefficients are strongly constrained to the $\mathcal{O}(10)$ range, resulting in a hierarchical flavor structure arising from a FN-like residual symmetry. 
In contrast, the relatively large ratios among coefficients in this study increase the likelihood of spontaneous CP violation, even in regions where $\Im\,[\tau]$ is relatively small compared with the results in Ref.~\cite{Abe:2023qmr}. 
However, note that we consider real coefficients for the mass matrices in Eq.~\eqref{eq:mass_matrix_up} and Eq.~\eqref{eq:mass_matrix_down}, whereas Ref.~\cite{Abe:2023qmr} explores complex coefficients. 
Thus, our results in the small $\Im\,[\tau]$ regime imply different properties compared to those found in previous studies.

\medskip

As mentioned above in relation to Fig.~\ref{fig:optimized_distribution}, $\Im\,[\tau]$ is strongly correlated with $\left\{ \alpha_2^1, \alpha_2^2, \alpha_3^2 \right\}$ but shows a weak correlation with $\left\{ \alpha_1, \beta_1, \beta_2, \beta_3^1, \beta_3^2 \right\}$. 
Based on this property, high-precision optimization is expected, where $\left\{ \alpha_1, \beta_1, \beta_2, \beta_3^1, \beta_3^2 \right\}$ can be adjusted without altering $\Im\,[\tau]$. 
This would allow the ratios among coefficients to be suppressed while maintaining the precision of $\chi^2$ fit.
In fact, among the solutions discussed in this paper that satisfy the criterion $\chi_{\mathrm{gen}} \lesssim 350$, there is an example in which the maximum ratio of coefficients prior to optimization reached 18.9. 
Given this case, it is entirely feasible to achieve coefficient ratios within $\mathcal{O}(10)$, similar to the previous study, without altering $\Im\,[\tau]$.
Verification of this point is left for future work.

\medskip

\medskip

\begin{table}[H]
\renewcommand{\arraystretch}{1.25}
\centering
\scalebox{0.88}{
   \begin{tabular}{|c||c|c|c|c|c|}\hline
	$\chi_{\mathrm{opt}}^2$ & $\mathbf{1.00}$ & $2.57$ & $3.54$ & $5.09$ & $6.97$ \\
        $\Re\,[\tau]$ & $\mathbf{-0.225}$ & $-0.414$ & $-0.363$ & $+0.345$ & $+0.533$ \\
        $\Im\,[\tau]$ & $\mathbf{2.24}$ & $2.25$ & $2.24$ & $2.25$ & $2.25$ \\
        $J/10^{-5}$ & $\mathbf{2.84}$ & $2.76$ & $2.82$ & $2.95$ & $2.85$ \\
        \hline
        $\chi_{\mathrm{opt}}^2$ & $7.15$ & $7.79$ & $8.42$ & $8.44$ & $8.50$ \\
	$\Re\,[\tau]$ & $-0.469$ & $+0.393$ & $+0.283$ & $-0.098$ & $+0.239$ \\
        $\Im\,[\tau]$ & $2.25$ & $2.26$ & $2.24$ & $2.25$ & $2.25$ \\
        $J/10^{-5}$ & $2.86$ & $2.79$ & $3.26$ & $2.75$ & $2.77$ \\
        \hline
    \end{tabular}
    }
  \caption{The values of modulus $\tau$ and Jarlskog invariant for the top 10 solutions with the smallest $\chi_{\mathrm{opt}}^2$ after the optimization for 611 solutions. The bold numbers correspond to Eq.~\eqref{eq:best_fit} indicating the best solution.}
    \label{tab:solutions}
\end{table}
\renewcommand{\arraystretch}{1}

\section{Conclusions}
\label{sec:con}

In order to understand the flavor structure of elementary particles, it is important to deepen our understanding of flavor models. 
Specifically, if we can find parameters that not only reproduce known observables but also reveal undiscovered properties, we can analyze the underlying structure of flavor models. 
In this study, we focused on the $S_4^\prime$ modular flavor model as a specific application of the diffusion model, a type of generative AI, and searched free parameters that reproduce experimental results. 
Conventional numerical methods typically require optimization by repeating the process to find out the desirable experimental values around fixed parameters as in the Monte-Carlo method. 
In contrast to those traditional approaches, the diffusion model provides a framework that is independent of the specific details of flavor models. 
Furthermore, it enables an inverse problem approach in which the machine provides a series of plausible model parameters from given experimental data. 

\medskip

In this paper, we constructed a diffusion model with MCMC to analyze the flavor structure of the quarks in the modular flavor model. 
Sec.~\ref{sec:DM_flavor} introduced the diffusion models and transfer learning in order to apply it to some flavor models. 
In Sec.~\ref{sec:S4prime_model}, the $S_4^\prime$ modular flavor model was used as a concrete example, and the diffusion model was applied to search for its free parameters. 
Following a brief review of the $S_4^\prime$ modular symmetry in Sec.~\ref{sec:S4prime_modular}, we organized the description of the quark sector of the $S_4^\prime$ model in Sec.~\ref{sec:S4prime_quark}. 
Specific representations and weights are selected based on the previous study to reproduce the semi-realistic flavor structure, so observables can be calculated by determining the free parameters. 
To optimize these parameters, the setup of our diffusion model is introduced in Sec.~\ref{sec:diffusion_model}, and physical implications derived from the data obtained by the diffusion model with MCMC are discussed in Sec.~\ref{sec:result}.

\medskip

Specifically, we confirmed that the $\chi^2$ value of the data generated by the diffusion model with MCMC is indeed improved, and exhibited how various parameter solutions are found as the number of data is increased. 
The generated results indicated that $\Im\,[\tau]$ was concentrated in a different region than in the previous study. 
Furthermore, it was found that the spontaneous CP violation appeared in the $S_4^\prime$ modular flavor model from $\Re\,[\tau]$. 
These findings demonstrate that the diffusion model can find semi-realistic parameters via experimental observables. 

\medskip

Before closing our paper, we will mention possible future works:

\begin{itemize}
    \item 
    Although our analysis considers only the quark sector, the lepton sector can also be analyzed by the same procedure. 
    Even within a framework that addresses both quarks and leptons simultaneously, there is no significant difference except that the inputs and outputs of the neural network have approximately double the dimensions. 
    Due to this characteristic, a search for free parameters in other modular flavor models can also be anticipated as a direct extension. 
    
    \item 
    The representation and modular weights of the fields used in this study were determined from an analytical perspective to achieve semi-realistic flavor structure and were fixed at specific values. 
    In the future, the application of machine learning could automate the selection of the expressions and weights themselves. 
    In fact, reinforcement learning, which is a type of machine learning, was utilized to search for $U(1)$ charges to assign to the fields in Refs.~\cite{Harvey:2021oue,Nishimura:2023nre,Nishimura:2024apb}. 
    By combining such techniques with diffusion models, it is possible to find out predictions of flavor models that have not been identified in previous empirical studies.
    
    \item 
    In light of the two aforementioned prospects, it is expected that various flavor models will be exhaustively explored by parameter searches based on the diffusion models, and one can compare the predictions of each model. 
    While this study involves 10 free parameters, an increase in the number of parameters is inevitable during dealing with many flavor models simultaneously. 
    On the other hand, state-of-the-art generative AIs can effectively manage vast parameter spaces, such as a $1{,}024 \times 1{,}024$ pixel image of high quality.\footnote{One of these technologies is SDXL by Stability AI \cite{Podell:2023sdxl}, which is improved version of Stable Diffusion.} 
    These technologies utilize large neural networks of course, but they also incorporate various efforts such as Variational Autoencoders (VAEs) and Transformers, which are also applied in particle physics. 
    By perceiving the set of free parameters as an image, a wide range of applications of the diffusion models in flavor physics becomes feasible.
\end{itemize}

\acknowledgments

This work was supported in part by Kyushu University’s Innovator Fellowship Program (S.N.), JSPS KAKENHI Grant Numbers JP25KJ1927 (S.N.), JP23H04512 (H.O.), JP25H01539  (H.O.), and JP26K07087 (H.O.).
This work forms part of the doctoral dissertation of Satsuki Nishimura at Kyushu University.

\bibliography{references}{}
\bibliographystyle{JHEP} 

\end{document}